\documentclass [11pt]{article}

\usepackage{doublespace}
\usepackage{epsfig}
\usepackage{amssymb}
\usepackage{verbatim}
\usepackage{delarray}
\usepackage{graphics}
\usepackage{epic}

\newcommand{\ket}[1]{{|#1\rangle}}
\newcommand{\bra}[1]{{\langle#1|}}
\newcommand{\braket}[2]{{\langle#1|#2\rangle}}

\newcommand{\phii}{$\ket{\phi_i} \,$}
\newcommand{\phiie}{\ket{\phi_i}}
\newcommand{\mui}{$\ket{\mu_i} \,$}
\newcommand{\muie}{\ket{\mu_i}}

\newcommand{\bl}{\left(}
\newcommand{\br}{\right)}

\newcommand{\tr}{\mbox{Tr}} 

\newcommand{\lra}{\leftrightarrow}
\newcommand{\inner}[2]{\langle{#1},{#2}\rangle}

\newcommand{\hme}{\hat{M}}
\newcommand{\hm}{$\hat{M} \,$}

\newcommand{\HH}{{\mathcal{H}}}
\newcommand{\V}{{\mathcal{V}}}
\newcommand{\U}{{\mathcal{U}}}
\newcommand{\C}{{\mathbb{C}}}
\newcommand{\Z}{{\mathbb{Z}}}
\newcommand{\G}{{\mathcal{G}}}
\newcommand{\SSS}{{\mathcal{S}}}
\newcommand{\FF}{{\mathcal{F}}}
\newcommand{\zerob}{{\mathbf 0}}

\newcommand{\ie}{{\em i.e., }}
\newcommand{\eg}{{\em e.g., }}

\newcommand{\etal}{\emph{et al.\ }}

\newtheorem{theorem}{Theorem}

\title{On Quantum Detection and the Square-Root Measurement}
\author{Yonina C. Eldar\footnote{Research Laboratory of Electronics,
Massachusetts Institute of Technology, Room 36-615,
Cambridge, MA 02139. E-mail: yonina@mit.edu.}
\hspace*{0.05in} and G. David Forney, Jr.\footnote{Laboratory for
Information and 
Decision Systems, 
Massachusetts Institute of Technology,
Cambridge, MA 02139. E-mail: forneyd@mediaone.net.}}

\date{\today}
 
\begin{document} 

\maketitle

\renewcommand{\thefootnote}{}
\footnotetext{This research was supported in part through
collaborative participation in the Advanced Sensors Consortium
sponsored by the U.S. Army Research Laboratory under Cooperative
Agreement DAAL01-96-2-0001 and supported in part by the Texas
Instruments Leadership University Program. Yonina Eldar is currently
supported by an IBM Research Fellowship.} 
\renewcommand{\thefootnote}{\arabic{footnote}}

%%%%%%%%%%%%%%%%%%%%%%%%%%%%%%%%%%%%%%%%%%%%%%%%%%%%%%%%%%
\begin{abstract}
In this paper we consider the problem of constructing 
measurements optimized to distinguish between a collection of possibly
non-orthogonal quantum states. 
We consider a collection of pure states and seek a 
positive operator-valued measure (POVM) consisting of
rank-one operators with measurement vectors closest in squared norm to
the given states. 
We compare our results to previous measurements suggested by 
Peres and Wootters \cite{PW91} and
Hausladen \etal \cite{Haus96}, where we  refer
to the latter as the square-root measurement (SRM). We obtain a new
characterization of the SRM, and prove that it is
optimal in a least-squares sense. In addition, we 
show that for a geometrically uniform state set the SRM minimizes 
the probability of a detection error.  This generalizes a similar
result of Ban \etal \cite{BKMO97}.
\end{abstract}
%%%%%%%%%%%%%%%%%%%%%%%%%%%%%%%%%%%%%%%%%%%%%%%%%%%%%%%%%%

%%%%%%%%%%%%%%%%%%%%%%%%%%
\section{Introduction}
\label{sec:intro}
%%%%%%%%%%%%%%%%%%%%%%%%%%

Suppose that a transmitter, Alice, wants to convey classical information to
a receiver, Bob, using a quantum-mechanical channel. Alice
represents  messages by preparing the quantum channel in a pure
quantum state drawn from a collection of known states.
Bob detects the information by subjecting the channel to a measurement
in order to determine the state prepared. 
If the quantum states are mutually orthogonal, 
then Bob can perform an optimal orthogonal (von Neumann) 
measurement that will determine the state correctly with
probability one \cite{Peresb}. The optimal measurement
consists of projections onto the given states.
However, if the given states are not orthogonal,
then no measurement will allow Bob to distinguish perfectly between them. 
Bob's problem is therefore to construct a
measurement optimized to distinguish between non-orthogonal pure
quantum states.

We may formulate this problem as a quantum detection problem, and  
seek a measurement that minimizes the
probability of a detection error, or more generally, minimizes the 
Bayes cost.
Necessary and sufficient conditions for an optimum measurement
minimizing the Bayes cost have been derived \cite{Hol73,Ken75,Helb}.
However, except in some particular cases \cite{Helb,Hel89,OBH96,BKMO97}, 
obtaining a closed-form analytical expression for the optimal
measurement directly from these conditions is 
a difficult and unsolved problem. Thus in practice, 
iterative procedures minimizing the Bayes cost
\cite{Hel82} or {\em ad-hoc} suboptimal measurements are used.

In this paper we take an alternative approach of choosing
a different optimality criterion, namely a squared-error criterion,
and seeking a measurement that
minimizes this criterion.
It turns out that the optimal measurement for this criterion is the
``square-root measurement'' (SRM), which has previously been proposed
as a ``pretty good'' {\em ad-hoc} measurement \cite{HW94,Haus96}. 

This work was originally motivated by the problems studied
by Peres and Wootters in \cite{PW91} and by Hausladen \etal in
\cite{Haus96}.
Peres and Wootters \cite{PW91} consider a source that emits three
two-qubit states with equal probability. In order to distinguish
between these states, they propose an orthogonal measurement consisting
of projections onto 
measurement vectors ``close'' to the given states. Their choice of measurement
results in a high probability of correctly determining the
state emitted by the source, and a large mutual information between the
state and the measurement outcome. 
However, they 
do not explain how they construct their measurement, and 
do not prove that it is optimal in any sense. 
Moreover, the measurement they propose is 
specific for the problem that they pose; they do not describe a
general procedure for constructing an orthogonal measurement with
measurement vectors close to given states. They also remark that improved
probabilities might be obtained by considering a general
positive operator-valued measure (POVM) \cite{Peres90} consisting of
positive Hermitian operators $\Pi_i$ satisfying $\sum_i\Pi_i=I$, where
the operators $\Pi_i$  are not required to be orthogonal projection
operators as in an orthogonal measurement. 

Hausladen \etal \cite{Haus96} consider the general problem of
distinguishing between an arbitrary set of pure states, where 
the number of states is no larger than the dimension of the space
$\U$ they span.
They describe a procedure for constructing a
general ``decoding observable'', corresponding to a POVM
consisting of rank-one operators
that distinguishes between the states ``pretty well''; this
measurement has subsequently been called the {\em square-root
measurement (SRM)} \mbox{(see \eg \cite{SKIH98,SUIH98,KOSH99})}. 
However, they make no assertion of (non-asymptotic) 
optimality. Although they mention the problem studied by Peres and
Wootters in \cite{PW91}, they make no connection between their
measurement and the Peres-Wootters measurement.

The SRM \cite{BKMO97,HW94,Haus96,SKIH98,SUIH98,KOSH99} has many desirable
properties. Its construction is relatively 
simple; it can be determined directly
from the given collection of states; it minimizes the probability
of a detection error when the states exhibit certain symmetries
\cite{BKMO97}; it is ``pretty good''  when the states to be
distinguished are equally likely and almost orthogonal \cite{HW94};
and it is asymptotically optimal \cite{Haus96}.
Because of these properties, the SRM has been employed as a detection
measurement in many applications (see \eg
\cite{SKIH98,SUIH98,KOSH99}). 
However, apart from some particular cases  mentioned
above \cite{BKMO97}, no assertion 
of (non-asymptotic) optimality is known for the SRM.

In this paper we systematically construct detection measurements 
optimized to distinguish between a collection of quantum states.
Motivated by the example studied by Peres and Wootters \cite{PW91}, we
consider pure-state 
ensembles and seek a POVM consisting
of rank-one positive operators with measurement vectors that minimize the  
sum of the squared norms of the error vectors, where the $i$th
error vector is defined as the difference between the $i$th
state vector and the $i$th measurement vector.
We refer to the optimizing measurement as the least-squares
measurement (LSM). 
We then generalize this approach to allow for unequal weighting of the 
squared norms of the error vectors. This weighted criterion may be of
interest when the 
given states have unequal prior probabilities. We refer to the
resulting measurement as the weighted least-squares measurement (WLSM).
We show that the SRM coincides with the LSM when
the prior probabilities are equal, and with the WLSM
otherwise (if the weights are proportional
to the square roots of the prior probabilities).   

We then consider the case in which the collection of states
has a strong symmetry property called geometric uniformity
\cite{F91}. We show that for such a state set the SRM minimizes the
probability of a detection error. This generalizes a similar 
result of Ban \etal \cite{BKMO97}.

The organization of this paper is as follows. 
In Section \ref{sec:prob} we formulate our problem and present our
main results. In Section \ref{sec:LSM} 
we construct a  measurement consisting of rank-one operators with
measurement vectors closest to a given collection of states in the least-squares
sense. In Section \ref{sec:CLSM} 
we construct the optimal orthogonal LSM.
Section \ref{sec:WLSM} generalizes these results to allow
for weighting of the squared norms of the error vectors.
In Section \ref{sec:compare}
we discuss the relationships between our results and the previous
results of Peres and Wootters \cite{PW91} and Hausladen \etal \cite{Haus96}. 
We obtain a new characterization of the SRM, and 
summarize the properties of the SRM that follow from this
characterization.
In Section \ref{sec:minpe} we discuss connections between the SRM
and the measurement minimizing the probability of a detection error
(MPEM). We show that for a geometrically uniform state set the SRM is
equivalent to the MPEM.
We will consistently use \cite{Haus96} as our principal
reference on the SRM.  

\newpage
%%%%%%%%%%%%%%%%%%%%%%%%%%%%%%%%%%%%%%%%%%%%%%%
\section{Problem Statement and Main Results} 
\label{sec:prob} 
%%%%%%%%%%%%%%%%%%%%%%%%%%%%%%%%%%%%%%%%%%%%%%%

In this section, we formulate our problem and describe our main results.

%%%%%%%%%%%%%%%%%%%%%%%%%%%%%%%%%%%
\subsection{Problem Formulation}
%%%%%%%%%%%%%%%%%%%%%%%%%%%%%%%%%%

Assume that Alice conveys classical information to Bob 
by preparing a quantum channel in a pure
quantum state drawn from a collection of given states $\{ \phiie \}$.
Bob's problem is to construct a measurement that will correctly determine the
state of the channel with high probability. 

Therefore, let $\{ \phiie \}$ be  a collection of $m\leq n$ normalized
vectors \phii in an
$n$-dimensional complex Hilbert space $\HH$. In general these vectors are 
non-orthogonal and span an \mbox{$r$-dimensional} subspace $\U
\subseteq \HH$. The vectors
are linearly independent if $r=m$.

For our measurement, we restrict our attention to POVMs consisting
of $m$ rank-one operators of the form
$\Pi_i=\muie \bra{\mu_i}$ with measurement vectors $\muie \in \U$. 
We do not require the vectors \mui to be orthogonal or
normalized. However, to constitute a POVM the measurement vectors must satisfy 
\begin{equation}
\label{eq:identu}
\sum_{i=1}^m \Pi_i =\sum_{i=1}^m  \muie \bra{\mu_i}=P_{\U},
\end{equation}
where $P_\U$ is the projection operator onto $\U$;  \ie the
operators $\Pi_i$ must be a resolution of the identity on
$\U$.\footnote{Often these operators are supplemented by a projection
$\Pi_0 = P_{\U^\perp} = I_{\HH} - P_{\U}$ onto the orthogonal subspace
$\U^\perp \subseteq \HH$, so that $\sum_{i=0}^m \Pi_i = I_{\HH}$--- \ie
the augmented POVM is a resolution of the identity on $\HH$.  However, if
the state vectors are confined to $\U$, then the probability of this
additional outcome is $0$, so we omit it.}

We seek the measurement vectors \mui such that one of
the following quantities is minimized:
\begin{enumerate}
\item \label{ls} Squared error $E = \sum_{i=1}^m \braket{e_i}{e_i}$, where
$\ket{e_i} = \phiie -\muie$;
\item \label{wls} Weighted squared error $E_w = \sum_{i=1}^m
 w_i\braket{e_i}{e_i}$  for a given set of positive weights $w_i$.
\end{enumerate}

%%%%%%%%%%%%%%%%%%%%%%%%%%%%%%%%%%%
\subsection{Main Results}
%%%%%%%%%%%%%%%%%%%%%%%%%%%%%%%%%%

If the states \phii are linearly independent (\ie if $r=m$), 
then the optimal solutions to problems (\ref{ls}) and (\ref{wls}) are
of the same general form. 
We express this optimal solution in different ways.  
In particular, we find that the optimal solution is an orthogonal 
measurement and not a general POVM.

If $r <m$, then the solution to problem (\ref{ls}) still has the same
general form. We show how it can be realized as an orthogonal
measurement in an $m$-dimensional space. This
orthogonal measurement is just a realization of
the optimal POVM in a larger space than $\U$, along the lines
suggested by Neumark's theorem \cite{Peres90},
and it furnishes a physical interpretation of the optimal POVM. 

We define a geometrically uniform (GU) state set as a  
collection of vectors $\SSS = \{\ket{\phi_i} =
U_i\ket{\phi}, U_i \in \G\}$, where
$\G$ is a finite abelian (commutative) group of $m$ unitary matrices
$U_i$, and
$\ket{\phi}$ is an arbitrary state. 
We show that for such a state set the SRM minimizes
the probability of a detection error.

Using these results, we can make the following remarks about
\cite{PW91} and the SRM \cite{Haus96}:
\begin{enumerate}
\item  The Peres-Wootters measurement is optimal in the
least-squares sense and is equal to the SRM  
(strangely, this was not noticed in \cite{Haus96}); it also
minimizes the probability of a detection error. 
\item  The SRM proposed by Hausladen \etal \cite{Haus96} minimizes the squared
error. It may always be chosen as an orthogonal measurement equivalent
to the optimal measurement in the linearly independent case. 
Further properties of the SRM are summarized in Theorem \ref{thm:srm} (Section
\ref{sec:compare}).  
\end{enumerate}

\newpage
%%%%%%%%%%%%%%%%%%%%%%%%%%%%%%%%%%%%%%%%%%%%%%%
\section{Least-Squares Measurement} 
\label{sec:LSM} 
%%%%%%%%%%%%%%%%%%%%%%%%%%%%%%%%%%%%%%%%%%%%%%%

Our  objective is to construct a POVM with measurement vectors $\ket{\mu_i}$,
optimized to distinguish 
between a collection of $m$ pure states \phii that span a space
$\U \subseteq \HH$.
A reasonable approach is to find a set of vectors $\muie \in \U$ 
that are ``closest'' to the  states \phii in the least-squares sense. 
Thus our measurement consists of $m$ rank-one 
positive operators of the form \mbox{$\Pi_i=\muie \bra{\mu_i}, 1 \leq
i \leq m$}.
The measurement vectors \mui are chosen to minimize the squared error $E$, defined by
\begin{equation}
\label{eq:serror}
E=\sum_{i=1}^m \braket{e_i}{e_i},
\end{equation} 
where $\ket{e_i}$ denotes the $i$th error vector
\begin{equation}
\label{eq:error}
\ket{e_i}=\phiie-\muie,
\end{equation} 
subject to the constraint (\ref{eq:identu}); \ie the operators $\Pi_i$
must be a resolution of the identity on $\U$.

If the vectors \phii are mutually orthonormal, then the solution to
(\ref{eq:serror}) satisfying the constraint (\ref{eq:identu}) is simply 
\mbox{$\phiie=\muie,\,\,1 \leq i\leq m$}, which yields $E=0$.

To derive the solution in the general case
where the vectors \phii are not orthonormal,
denote by $M$ and $\Phi$ the $n \times m$ matrices whose columns are the
vectors \mui and $\ket{\phi_i}$, respectively. The 
squared error $E$ of (\ref{eq:serror})-(\ref{eq:error}) may then be
expressed in terms of these matrices  as 
\begin{equation}
\label{eq:errorm}
E=\tr\bl (\Phi-M)^*(\Phi-M) \br=\tr\bl (\Phi-M)(\Phi-M)^* \br,
\end{equation} 
where $\tr(\cdot)$ and $(\cdot)^*$ denote the trace and the Hermitian
conjugate respectively, and the second equality follows from 
the identity $\tr(AB)=\tr(BA)$ for all matrices $A, B$.
The constraint (\ref{eq:identu}) may then be restated as
\begin{equation}
\label{eq:constm}
MM^*=P_{\U}.
\end{equation}

%%%%%%%%%%%%%%%%%%%%%%%%%%%%%%%%%%%%%%%%%%%%%%%%%%
\subsection{The Singular Value Decomposition}
\label{sec:svd}
%%%%%%%%%%%%%%%%%%%%%%%%%%%%%%%%%%%%%%%%%%%%%%%%%%
The least-squares problem of (\ref{eq:errorm}) seeks a measurement matrix $M$
that is ``close'' to the matrix $\Phi$. If the two matrices are close,
then we expect that the underlying linear transformations
they represent will share similar properties.
We therefore begin by decomposing the matrix $\Phi$ 
into elementary matrices that reveal these properties
via the {\em singular value decomposition} \mbox{(SVD)
\cite{GVL}}.

The SVD is known in quantum mechanics, but possibly not
very well known.  It has sometimes been presented as a corollary of the
polar decomposition (\eg in Appendix A of \cite{KR00}).  We present here a
brief derivation based on the properties of eigendecompositions, since the
SVD can be interpreted as a sort of ``square root'' of an eigendecomposition.

Let $\Phi$ be an arbitrary  $n \times m$ complex matrix of rank $r$.
Theorem \ref{thm:svd} below
asserts that $\Phi$ has a SVD of the form
$\Phi = U\Sigma V^*$, 
with $U$ and $V$ unitary matrices and $\Sigma$
diagonal. The elements of the SVD
may be found from the eigenvalues and eigenvectors of
the $m \times m$ non-negative definite Hermitian matrix $S =\Phi^*\Phi$ and
the $n \times n$ non-negative definite Hermitian matrix $T =\Phi\Phi^*$.  
Notice that $S$ is the Gram matrix of inner products
$\braket{\phi_i}{\phi_j}$, which completely determines the relative geometry
of the vectors $\{\ket{\phi_i}\}$.
It is elementary that both $S$ and $T$ have the same rank
$r$ as $\Phi$, and that their nonzero eigenvalues are the same set of $r$
positive numbers $\{\sigma_i^2, 1 \leq i \leq r\}$.

\begin{theorem}[Singular Value Decomposition (SVD)]
\label{thm:svd}
Let $\{\ket{\phi_i}\}$ be a set of $m$ vectors in an
\mbox{$n$-dimensional} complex 
Hilbert space $\HH$, let $\U \subseteq \HH$ be the subspace spanned by these
vectors, and let $r = \dim~\U$.  Let $\Phi$ be the rank-$r$ $n \times m$
matrix whose columns are the vectors $\{\ket{\phi_i}\}$.  Then
\[ \Phi = U\Sigma V^* = \sum_{i=1}^r \sigma_i\ket{u_i}\bra{v_i}, \]
where
\begin{enumerate}
\item $\Phi^*\Phi = V(\Sigma^*\Sigma) V^* = \sum_{i=1}^r
\sigma_i^2\ket{v_i}\bra{v_i}$ is an eigendecomposition of the rank-$r$ $m
\times m$ matrix $S = \Phi^*\Phi$,
in which
\begin{enumerate}
\item the $r$ positive real numbers $\{\sigma_i^2, 1 \leq i \leq r\}$ are the
nonzero eigenvalues of $S$, and $\sigma_i$ is the positive square root of
$\sigma_i^2$;
\item the $r$ vectors $\{\ket{v_i} \in \,\,\C^m, 1 \leq i \leq r\}$
are the corresponding eigenvectors in the $m$-dimensional complex
Hilbert space $\C^m$, normalized so that
$\braket{v_i}{v_i} = 1$; 
\item $\Sigma$ is a diagonal $n \times m$ matrix whose first $r$
diagonal elements are $\sigma_i$, and whose remaining $m-r$
diagonal elements are $0$, so $\Sigma^*\Sigma$ is a diagonal $m \times m$
matrix with diagonal elements $\sigma_i^2$ for $1 \leq i \leq r$ and $0$
otherwise; 
\item $V$ is an $m \times m$ unitary matrix whose first $r$ columns are the
eigenvectors $\ket{v_i}$, which span a subspace $\V \subseteq \C^m$, and
whose remaining $m-r$ columns $\ket{v_i}$ span the orthogonal
complement $\V^\perp \subseteq \C^m$;
\end{enumerate}
and
\item $\Phi\Phi^* = U(\Sigma\Sigma^*) U^* = \sum_{i=1}^r
\sigma_i^2\ket{u_i}\bra{u_i}$ is an eigendecomposition of the rank-$r$
$n \times n$ matrix $T = \Phi\Phi^*$,
in which
\begin{enumerate}
\item the $r$ positive real numbers $\{\sigma_i^2, 1 \leq i \leq r\}$ are as
before, but are now identified as the nonzero eigenvalues of $T$;
\item the $r$ vectors $\{\ket{u_i} \in \,\,$$\HH$$, 1 \leq i \leq r\}$ are the
corresponding eigenvectors, normalized so that $\braket{u_i}{u_i} = 1$;
\item $\Sigma$ is as before, so $\Sigma\Sigma^*$ is a diagonal $n \times n$
matrix with diagonal elements $\sigma_i^2$ for $1 \leq i \leq r$ and $0$
otherwise;
\item $U$ is an $n \times n$ unitary matrix whose first $r$ columns are the
eigenvectors $\ket{u_i}$, which span the subspace $\U \subseteq \HH$, and
whose remaining $n-r$ columns $\ket{u_i}$ span the orthogonal
complement $\U^\perp \subseteq \HH$.
\end{enumerate}
\end{enumerate}
\end{theorem}

Since $U$ is unitary, we have not only $U^*U = I_\HH$, which implies that
the vectors $\ket{u_k} \in \HH$ are orthonormal,
$\braket{u_k}{u_j} = \delta_{kj}$, but also that $UU^* = I_\HH$, which
implies that the rank-one projection operators $\ket{u_k}\bra{u_k}$ are a
resolution of the identity, $\sum_k \ket{u_k}\bra{u_k} = I_\HH$.
Similarly the vectors $\ket{v_k} \in \C^m$ are orthonormal and
$\sum_k \ket{v_k}\bra{v_k} = I_m$.
These orthonormal bases for $\HH$ and $\C^m$ will be called the
$U$-basis and the $V$-basis, respectively. The first $r$ vectors of
the $U$-basis and the $V$-basis span the subspaces $\U$ and $\V$,
respectively. Thus we refer to the set of vectors
$\{\ket{u_k}, 1 \leq k \leq r\}$ as the $\U$-basis, and to the set
$\{\ket{v_k}, 1 \leq
k \leq r\}$ as the $\V$-basis.

The matrix $\Phi$ may be viewed as defining a linear transformation
$\Phi:\C^m \to \HH$
according to $\ket{v} \mapsto \Phi\ket{v}$.  The
SVD allows us to interpret this map as follows.
A vector $\ket{v} \in \C^m$
is first decomposed into its $V$-basis components via $\ket{v} = \sum_i
\ket{v_i}\braket{v_i}{v}$.  Since $\Phi$ maps $\ket{v_i}$ to
$\sigma_i\ket{u_i}$, $\Phi$ maps the $i$th component
$\ket{v_i}\braket{v_i}{v}$ to $\sigma_i\ket{u_i}\braket{v_i}{v}$.
Therefore, by superposition, $\Phi$ maps $\ket{v}$ to $\sum_i
\sigma_i\ket{u_i}\braket{v_i}{v}$.  The kernel of the map $\Phi$ is thus
$\V^\perp \subseteq \C^m$, and its image is $\U \subseteq \HH$.

\begin{sloppypar}
Similarly, the conjugate Hermitian matrix $\Phi^*$ defines the adjoint
linear transformation \mbox{$\Phi^*:\HH \to \C^m$} as follows:
$\Phi^*$ maps $\ket{u} \in \HH$ to $\sum_i
\sigma_i\ket{v_i}\braket{u_i}{u} \in \C^m$.  The 
kernel of the adjoint map $\Phi^*$ is thus $\U^\perp \subseteq \HH$,
and its image is $\V \subseteq \C^m$.
\end{sloppypar}

The key element in these maps is the ``transjector" (partial isometry)
$\ket{u_i}\bra{v_i}$,
which maps the rank-one eigenspace of $S$ generated by $\ket{v_i}$ into the
corresponding eigenspace of $T$ generated by $\ket{u_i}$, and the adjoint
transjector $\ket{v_i}\bra{u_i}$, which performs the inverse map.

%%%%%%%%%%%%%%%%%%%%%%%%%%%%%%%%%%%%%%%%%%%%%%%%%%%%%%%%%%%
\subsection{The Least-Squares POVM}
%%%%%%%%%%%%%%%%%%%%%%%%%%%%%%%%%%%%%%%%%%%%%%%%%%%%%%%%%%%%

The SVD of $\Phi$ specifies orthonormal bases for
$\V$ and $\U$ such that the linear transformations 
$\Phi$ and $\Phi^*$ map one basis to the other with appropriate scale
factors. Thus, to find an $M$ close to $\Phi$ we need to find a linear
transformation $M$ that performs a map similar to $\Phi$.

Employing the SVD $\Phi=U\Sigma V^*$, 
we rewrite the squared error $E$ of (\ref{eq:errorm}) as
\begin{equation}
\label{eq:errorm2}
E =\tr\bl (\Phi-M)(\Phi-M)^*\br=\tr\bl
U^*(\Phi-M)(\Phi-M)^*U\br=\sum_{i=1}^n \braket{d_i}{d_i},
\end{equation} 
where
\begin{equation}
\ket{d_i}=(\Phi-M)^*\ket{u_i}.
\end{equation}

The vectors $\{\ket{u_i},\,1 \leq i \leq r\}$ form an orthonormal
basis for $\U$. 
Therefore, the projection operator onto $\U$ is given by
\begin{equation} 
\label{eq:identusvd}
P_{\U} =\sum_{i=1}^r \ket{u_i} \bra{u_i}.
\end{equation}

Essentially, we want to construct a map $M^*$ such that the images
of the maps defined by $\Phi^*$ and $M^*$  are as close as
possible in the squared norm sense, subject to the  constraint
\begin{equation} 
\label{eq:constmsvd}
MM^* =\sum_{i=1}^r \ket{u_i} \bra{u_i}.
\end{equation}

The SVD of $\Phi^*$ is given by $\Phi^*=V \Sigma^* U^*$. Consequently,
\begin{equation}
\label{eq:lspui}
\Phi^*\ket{u_i}=
\left\{ 
\begin{array}{ll}
\sigma_i \ket{v_i}, \hspace{0.1in} & 1 \leq i \leq r; \nonumber \\
\ket{0}, & r+1 \leq i \leq n,
\end{array}
\right.
\end{equation}
where $\ket{0}$ denotes the zero vector.
Denoting the image of $\ket{u_i}$ under 
$M^*$ by $\ket{a_i}=M^* \ket{u_i}$, for any choice of $M$ satisfying the
constraint (\ref{eq:constmsvd}) we have
\begin{equation}
\label{eq:lsmui}
\braket{a_i}{a_i}=\bra{u_i} MM^* \ket{u_i}=
\left\{
\begin{array}{ll}
1, \hspace{0.1in}& 1 \leq i \leq r; \nonumber \\
0, & r+1 \leq i \leq n, 
\end{array}
\right.
\end{equation}
and 
\begin{equation}
\braket{a_i}{a_j}=\bra{u_i} MM^* \ket{u_j}=0, \,\,i \neq j.
\end{equation}
Thus the vectors $\ket{a_i},\,\,1 \leq i \leq r$, are mutually
orthonormal and $\ket{a_i}=\ket{0},\,\,r+1 \leq i \leq n$.
Combining (\ref{eq:lspui}) and (\ref{eq:lsmui}),
we may express $\ket{d_i}$ as
\begin{equation}
\ket{d_i}=
\left\{
\begin{array}{ll}
\sigma_i \ket{v_i}-\ket{a_i}, \hspace{0.1in}& 1 \leq i \leq r; \nonumber \\
\ket{0}, & r+1 \leq i \leq n.
\end{array}
\right.
\end{equation}

Our problem therefore reduces to finding
a set of $r$ orthonormal vectors $\ket{a_i}$ that minimize
$E=\sum_{i=1}^r \braket{d_i}{d_i}$, where $\ket{d_i}=\sigma_i
\ket{v_i}-\ket{a_i}$. Since the vectors
$\ket{v_i}$ are 
orthonormal, the minimizing vectors must be \mbox{$\ket{a_i}=
\ket{v_i},\,\,1 \leq i\leq r$}. 

Thus the optimal measurement matrix $M$, denoted
by \hm, satisfies
\begin{equation}
\hme^* \ket{u_i}=
\left\{
\begin{array}{ll}
\ket{v_i}, \hspace{0.1in}& 1 \leq i \leq r; \nonumber \\
\ket{0}, & r+1 \leq i \leq n.
\end{array}
\right.
\end{equation}
Consequently
\begin{equation}
\label{eq:lsm}
\hme=\sum_{i=1}^r\ket{u_i} \bra{v_i}.
\end{equation} 
In other words, the optimal \hm is just the sum of the $r$ transjectors of the
map $\Phi$. \\
We may express \hm in matrix form as
\begin{equation}
\label{eq:lsmm}
\hme=UZ_rV^*,
\end{equation}
where $Z_r,\,1 \leq r \leq m$ is an $n \times m$ matrix defined by 
\begin{equation}
\label{eq:zr}
Z_r=\left[
\begin{array}{c|c}
I_r & 0  \\
\hline
0&0 \\
\end{array} \right].
\end{equation} 

The residual squared error is then
\begin{equation}
\label{eq:emin}
E_{min}=\sum_{i=1}^r (1-\sigma_i)^2 \braket{v_i}{v_i}=\sum_{i=1}^r
(1-\sigma_i)^2.
\end{equation}
Recall that $S=\Phi^*\Phi=V \Sigma^* \Sigma V^*$; thus $\tr(S)=\sum_{i=1}^r
\sigma_i^2$. Also, if the vectors \phii are normalized, then 
the diagonal elements
of $S$ are all equal to $1$, so $\tr(S)= m$. Therefore,
\begin{equation}
\label{eq:emin2}
E_{min}=\sum_{i=1}^r (1-\sigma_i)^2=r+m-2\sum_{i=1}^r \sigma_i.
\end{equation}

Note that if the singular values $\sigma_i$ are distinct, then the
vectors $\ket{u_i},\,\,1 \leq i \leq r$ are unique (up to a phase
factor $e^{j\theta_i}$). Given the vectors $\ket{u_i}$, the vectors
$\ket{v_i}$ are uniquely determined, so the optimal measurement vectors
corresponding to \hm are unique.  

If on the other hand 
there are repeated singular values, then the 
corresponding eigenvectors are not unique. Nonetheless, the
choice of basis does not affect $\hat{M}$. Indeed,
if the eigenvectors corresponding to a repeated eigenvalue are
$\{\ket{u_j}\}$, then $\sum_j \ket{u_j} \bra{u_j}$ is a projection onto
the corresponding eigenspace, and therefore is the same regardless of
the choice of the eigenvectors
$\{\ket{u_j}\}$. Thus $\sum_j \ket{u_j}\bra{v_j} =\sum_j
\ket{u_j}\bra{u_j}\Phi$, independent of the choice of $\{\ket{u_j}\}$, and
the optimal measurement is unique.

We may express \hm directly in terms of $\Phi$ as
\begin{equation}
\label{eq:lsmphi}
\hme=\Phi((\Phi^*\Phi)^{1/2})^{\dagger},
\end{equation}
where $(\cdot)^\dagger$ denotes the {\em Moore-Penrose pseudo-inverse}
\cite{GVL}; the inverse is taken on the subspace
spanned by the columns of the matrix. Thus
\mbox{$((\Phi^* \Phi)^{1/2})^\dagger=V((\Sigma^*
\Sigma)^{1/2})^{\dagger} V^*$},
where $((\Sigma^* \Sigma)^{1/2})^\dagger$ is a diagonal matrix with diagonal
elements $1/\sigma_i$ for $1 \leq i\leq r$ and $0$ otherwise;
consequently, $\Phi((\Phi^*\Phi)^{1/2})^{\dagger}=UZ_rV^*$.

Alternatively, \hm may be expressed as
\begin{equation}
\label{eq:lsmphi2}
\hme=((\Phi\Phi^*)^{1/2})^ \dagger\Phi,
\end{equation}
where 
\mbox{$((\Phi\Phi^*)^{1/2})^\dagger=U((\Sigma
\Sigma^*)^{1/2})^{\dagger} U^*$}.
In Section \ref{sec:compare} we will
show that (\ref{eq:lsmphi2}) is equivalent to the SRM
proposed by Hausladen \etal \cite{Haus96}. 

In Appendix A we discuss some of the properties of the
residual squared error $E_{min}$.

%%%%%%%%%%%%%%%%%%%%%%%%%%%%%%%%%%%
\section{Orthogonal Least-Squares Measurement}
\label{sec:CLSM}
%%%%%%%%%%%%%%%%%%%%%%%%%%%%%%%%%%

In the previous section we sought the POVM consisting of
rank-one operators that minimizes the least-squares error.
We may similarly seek the optimal orthogonal measurement of the same form.
We will explore the connection between the resulting optimal
measurements both in the case of linearly independent states \phii ($r=m$),
and in the case of linearly dependent states ($r<m$).

{\em Linearly independent states}:
If the states \phii are linearly independent and consequently  $\Phi$ has
full column rank (\ie $r=m$), then (\ref{eq:lsmphi}) reduces to
\begin{equation}
\label{eq:lsmphired}
\hme=\Phi(\Phi^*\Phi)^{-1/2}.
\end{equation}
The optimal measurement vectors $\ket{\hat{\mu}_i}$ are mutually orthonormal, since 
their Gram matrix is
\begin{equation}
\hme^*\hme=(\Phi^*\Phi)^{-1/2}\Phi^*\Phi(\Phi^*\Phi)^{-1/2}=I_m.
\end{equation}
Thus, the optimal POVM is in fact an orthogonal measurement corresponding
to projections onto a set of mutually orthonormal measurement vectors, 
which must of course be the optimal orthogonal measurement as well.

{\em Linearly dependent states}:
If the vectors \phii are linearly dependent, so that the matrix $\Phi$ does not
have full column rank (\ie $r < m$), then the $m$ measurement vectors
$\ket{\hat{\mu}_i}$ cannot be 
mutually orthonormal since they span an $r$-dimensional subspace.
We therefore seek the orthogonal measurement $M$ that minimizes the
squared error $E$ 
given by (\ref{eq:errorm}),
subject to the orthonormality constraint \mbox{$M^*M=I_m$}.

In the previous section the constraint was on $MM^*$. Here
the constraint is on $M^*M$, so we  
now write the squared error $E$ as: 
\begin{equation}
\label{eq:vnserror}
E=\tr\bl (\Phi-M)^*(\Phi-M) \br
=\tr\bl V^*(\Phi-M)^*(\Phi-M) V\br=
\sum_{i=1}^m \braket{\tilde{d}_i}{\tilde{d}_i}, 
\end{equation} 
where
\begin{equation}
\ket{\tilde{d}_i}=(\Phi-M)\ket{v_i},
\end{equation}
and where the columns $\ket{v_i}$ of $V$ form the $V$-basis in the SVD
of $\Phi$. 
Essentially, we now want the images of the maps defined by $\Phi$ and $M$ 
to be as close as possible in the squared norm sense.

The SVD of $\Phi$ is given by $\Phi=U \Sigma V^*$. Thus,
\begin{equation}
\label{eq:vnpui}
\Phi \ket{v_i}=\left\{
\begin{array}{ll}
\sigma_i \ket{u_i}, \hspace{0.1in} & 1 \leq i \leq r; \nonumber \\
\ket{0}, & r+1 \leq i \leq m.
\end{array}
\right.
\end{equation}
Denoting the images of $\ket{v_i}$ under 
$M$ by $\ket{b_i}=M \ket{v_i}$, it follows from the constraint
$M^*M=I$ that the vectors
\mbox{$\ket{b_i},\,\,1 \leq i \leq m$}, are orthonormal. 

Our problem therefore reduces to finding
a set of $r$ orthonormal vectors $\ket{b_i}$ that minimize
$\sum_{i=1}^r \braket{\tilde{d_i}}{\tilde{d_i}}$, where
$\ket{\tilde{d_i}}= \sigma_i \ket{u_i}-\ket{b_i}$ (since
$\sum_{i=r+1}^m \braket{\tilde{d_i}}{\tilde{d_i}}= \sum_{i=r+1}^m
\braket{b_i}{b_i}=m-r$   
independent of the choice of \mbox{$\ket{b_i},\,\,r+1 \leq i \leq m$}). 
Since the vectors $\ket{u_i}$ are orthonormal, the minimizing vectors
must be $\ket{b_i}=\ket{u_i},\,\,1 \leq i\leq r$. 

We may choose the remaining vectors 
$\ket{b_i},\,\,r+1 \leq i\leq m$, arbitrarily, as long as the
resulting $m$ vectors $\ket{b_i}$ are mutually orthonormal. This
choice will not affect the 
residual squared error.
A convenient choice 
is $\ket{b_i}=\ket{u_i},\,\,r+1 \leq i\leq m$. This results in an
optimal measurement matrix denoted by $\tilde{M}$, namely
\begin{equation}
\label{eq:mperp}
\tilde{M}=\sum_{i=1}^m\ket{u_i} \bra{v_i}.
\end{equation} 
We may express $\tilde{M}$ in matrix form as
\begin{equation}
\label{eq:mperpm}
\tilde{M}=UZ_mV^*,
\end{equation} 
where $Z_m$ is given by (\ref{eq:zr}) with $r=m$.

The residual squared error is then
\begin{equation}
\label{eq:errvn}
\tilde{E}_{min}=\sum_{i=1}^r (1-\sigma_i)^2 \braket{u_i}{u_i}+\sum_{i=r+1}^m
\braket{u_i}{u_i}=\sum_{i=1}^r(1-\sigma_i)^2+m-r=E_{min}+m-r,
\end{equation}
where $E_{min}$ is given by (\ref{eq:emin}).

Evidently, the optimal orthogonal measurement is not strictly
unique. However, its action in the subspace $\U$ spanned by the vectors
\phii and the resulting $\tilde{E}_{min}$ are unique.

%%%%%%%%%%%%%%%%%%%%%%%%%%%%%%%%%%%%%%%%%%%%%%%%%%%%%%%%%%%%%%%%
\subsection{The Optimal Measurement and Neumark's Theorem}
%%%%%%%%%%%%%%%%%%%%%%%%%%%%%%%%%%%%%%%%%%%%%%%%%%%%%%%%%%%%%%%%
We now try to gain some insight into  the orthogonal
measurement.
Our problem is to find a set of measurement vectors that are as close
as possible to the states \phii, where the states lie in an
$r$-dimensional subspace $\U$. When $r=m$ we
showed that the optimal measurement vectors $\ket{\hat{\mu}_i}$ are mutually
orthonormal. However, when
$r<m$, there are at most $r$ orthonormal vectors in
$\U$. Therefore,  imposing an orthogonality constraint 
forces the optimal orthonormal measurement vectors
$\ket{\tilde{\mu}_i}$ to lie partly in the orthogonal
complement $\U^\perp$.  The corresponding measurement
consists of projections onto $m$ orthonormal measurement vectors,
where each vector has 
a component in $\U$, $\ket{\tilde{\mu}_i^\U}$, and a component in
$\U^\perp$, $\ket{\tilde{\mu}_i^{\U^\perp}}$.  
We may express $\tilde{M}$  in terms of these components as
\begin{equation}
\label{eq:mupmu}
\tilde{M}=\tilde{M}^\U+\tilde{M}^{\U^\perp},
\end{equation} 
where $\ket{\tilde{\mu}_i^\U}$ and $\ket{\tilde{\mu}_i^{\U^\perp}}$
are the columns of $\tilde{M}^\U$ and $\tilde{M}^{\U^\perp}$, respectively. 
From (\ref{eq:mperp}) it then follows that
\begin{equation}
\label{eq:partu}
\tilde{M}^\U=\sum_{i=1}^r\ket{u_i} \bra{v_i},
\end{equation}  
and
\begin{equation}
\label{eq:partup}
\tilde{M}^{\U^\perp}=\sum_{i=r+1}^m\ket{u_i} \bra{v_i}.
\end{equation}  
Comparing (\ref{eq:partu}) with (\ref{eq:lsm}), we conclude that 
$\tilde{M}^\U=\hme$ and therefore
$\ket{\tilde{\mu}_i^\U}=\ket{\hat{\mu}_i}$. Thus, although
$\ket{\tilde{\mu}_i} \neq \ket{\hat{\mu}_i}$, their components in
$\U$ are equal; \ie $P_\U \ket{\tilde{\mu}_i} = \ket{\hat{\mu}_i}$.
 
Essentially, the optimal orthogonal measurement seeks $m$ orthonormal measurement vectors
$\ket{\tilde{\mu}_i}$ whose projections onto $\U$ are as close as possible to
the $m$ states \phii. We now see that these projections are
the measurement vectors $\ket{\hat{\mu}_i}$ of the optimal POVM. 
If we consider only the components of the measurement vectors that lie
in $\U$, then  
\mbox{$\tilde{E}_{min}=\sum_{i=1}^r (1-\sigma_i)^2\braket{u_i}{u_i}=E_{min}$}.

Indeed, Neumark's theorem \cite{Peres90} shows that 
 our optimal orthogonal measurement is just a realization of
the optimal POVM.
This theorem guarantees that any
POVM with measurement operators of the form $\Pi_i=\muie \bra{\mu_i}$ may be
 realized by a set of 
orthogonal projection operators $\tilde{\Pi}_i$ in an extended space
such that $\Pi_i=P \tilde{\Pi}_i P$,
where $P$ is the projection operator onto the original smaller space.
Denoting by $\hat{\Pi}_i$ and $\tilde{\Pi}_i$ the optimal rank-one operators
$\ket{\hat{\mu}_i} \bra{\hat{\mu}_i}$ and $\ket{\tilde{\mu}_i}
\bra{\tilde{\mu}_i}$ respectively, (\ref{eq:partu}) asserts that 
\begin{equation}
\hat{\Pi}_i=P_\U \tilde{\Pi}_i P_\U.
\end{equation}

Thus the optimal orthogonal measurement is a set of $m$ projection
operators in $\HH$ that realizes the optimal POVM in the
$r$-dimensional space $\U
\subseteq \HH$. This furnishes a physical interpretation of the
optimal POVM. 
The two measurements are equivalent on the \mbox{subspace $\U$}.

\newpage
We summarize our results regarding the LSM in the following theorem:
\begin{theorem}[Least-squares measurement (LSM)]
\label{thm:ls}
Let $\{ \ket{\phi_i} \}$ be a set of $m$ vectors in an
$n$-dimensional complex 
Hilbert space $\HH$ that span an $r$-dimensional subspace $\U
\subseteq \HH$.
Let $\{ \ket{\hat{\mu}_i} \}$ denote the optimal $m$ measurement
vectors that minimize the 
least-squares error defined by (\ref{eq:serror})-(\ref{eq:error}),
subject to the constraint (\ref{eq:identu}).
Let $\Phi=U\Sigma V^*$ be the
rank-$r$ $n \times m$ 
matrix whose columns are the vectors $\ket{\phi_i}$, and let \hm
be the $n \times m$ measurement matrix whose columns are the vectors
$\ket{\hat{\mu}_i}$. 
Then the unique optimal \hm is given by
\[ \hme=\sum_{i=1}^r\ket{u_i}\bra{v_i}=UZ_rV^*=
\Phi((\Phi^*\Phi)^{1/2})^{\dagger}=   
((\Phi\Phi^*)^{1/2})^ \dagger\Phi, \]
where $\ket{u_i}$ and $\ket{v_i}$ denote the columns of $U$ and $V$
respectively, and $Z_r$ is defined in (\ref{eq:zr}).\\
The residual squared error is given by
\[ E_{min}=\sum_{i=1}^r (1-\sigma_i)^2=r+m-2\sum_{i=1}^r \sigma_i, \]
where $\{ \sigma_i,\,1 \leq i \leq r \}$ are the nonzero singular
values of $\Phi$. 
In addition,
\begin{enumerate}
\item If $r=m$, 
\begin{enumerate}
\item $\hme=\Phi(\Phi^*\Phi)^{-1/2}$;
\item $\hme^*\hme=I_m$ and the corresponding measurement is an
orthogonal measurement.
\end{enumerate}
\item If $r<m$,
\begin{enumerate}
\item $\hme$ may be realized by the optimal orthogonal measurement 
\mbox{$\tilde{M}=\sum_{i=1}^m\ket{u_i} \bra{v_i}=UZ_mV^*$}; 
\item the action of the two optimal measurements in the subspace
$\U$ is the same. 
\end{enumerate}
\end{enumerate}
\end{theorem}

%%%%%%%%%%%%%%%%%%%%%%%%%
\section{Weighted Least-Squares Measurement}
\label{sec:WLSM}
%%%%%%%%%%%%%%%%%%%%%%%

In the previous section  we sought a set of 
vectors \mui to minimize the sum of the squared errors, $E=\sum_{i=1}^m
\braket{e_i}{e_i}$, where $\ket{e_i}=\phiie-\muie$ is the $i$th error vector.
Essentially, we are assigning
equal weights to the different errors.
However, in many cases we might choose to
weight these errors according to some prior
knowledge regarding the states $\ket{\phi_i}$. For example, if the state
$\ket{\phi_j}$ is prepared with high probability, then we might wish
to assign a large weight to $\braket{e_j}{e_j}$. It may therefore be
of interest to seek the vectors \mui that minimize a weighted squared error.

Thus we consider the more general problem of minimizing the
weighted squared error $E_w$ given by 
\begin{equation}
\label{eq:werror}
E_w=\sum_{i=1}^m w_i \braket{e_i}{e_i}=
\sum_{i=1}^m w_i (\bra{\phi_i}-\bra{\mu_i})(\phiie-\muie),
\end{equation} 
subject to the constraint
\begin{equation}
\sum_{i=1}^m \muie \bra{\mu_i}=P_{\U},
\end{equation}
where $w_i>0$ is the weight given to 
the $i$th squared norm error. Throughout this section we will assume that
the vectors \phii are linearly independent and normalized.

The derivation of the solution to this minimization problem is 
analogous to the derivation of the LSM with a slight
modification.
In addition to the the matrices $M$ and $\Phi$,
we define an $m \times m$  diagonal matrix $W$ with diagonal elements $w_i$.
We further define $\Phi_w=\Phi W$. 
We then express $E_w$ in terms of $M, \Phi_w$ and $W$ as 
\begin{eqnarray}
\label{eq:errorwm2}
E_w & = & \tr \bl (\Phi-M)^*(\Phi-M)W \br \nonumber \\
    & = & \tr\bl (\Phi_w-M)(\Phi_w-M)^*\br +\tr \bl (W-I_m)M^*M \br +
                     \tr \bl W(I_m-W)\Phi^*\Phi \br.   
\end{eqnarray} 

From (\ref{eq:identusvd}) and (\ref{eq:constmsvd}), $M$ must satisfy
$MM^*=\sum_{i=1}^m \ket{u_i} \bra{u_i}=P_{\U}$, 
where $\ket{u_i}$ are the columns of $U$, the 
$U$-basis in the SVD of $\Phi$.
Consequently, $M$ must be of the form
$M=\sum_{i=1}^m \ket{u_i} \bra{q_i}$, where the $\ket{q_i}$ are
orthonormal vectors in $\C^m$,
from which it follows that $M^*M=I_m$. 
Thus, $\tr \bl W(I_m-W)M^*M \br=\tr \bl W(I_m-W)\br$. Moreover, since
$W(I_m-W)$ is diagonal and the vectors \phii are normalized, we have
$\tr \bl W(I_m-W)\Phi^*\Phi \br=\tr \bl W(I_m-W)\br$.  
Thus we may express the squared error $E_w$ as
\begin{equation}
\label{eq:esum}
E_w=\tr\bl (\Phi_w-M)(\Phi_w-M)^*\br-\tr \bl (I_m-W)(I_m-W) \br=
E'_w-\sum_{i=1}^m(1-w_i)^2, 
\end{equation}
where $E'_w$ is defined as 
\begin{equation}
\label{eq:nwerr}
E'_w  = \tr\bl (\Phi_w-M)(\Phi_w-M)^*\br.
\end{equation} 

Thus minimization of $E_w$ is equivalent to minimization of
$E'_w$. Furthermore, 
this minimization problem is equivalent to the least-squares minimization
given by (\ref{eq:errorm}), if we substitute $\Phi_w$ for $\Phi$. 

Therefore we now employ
the SVD of $\Phi_w$, namely $\Phi_w=U_w \Sigma_w V_w^*$. Since $W$ is
assumed to be invertible, the space spanned
by the columns of $\Phi_w=\Phi W$ is equivalent to the space spanned by the
columns of $\Phi$, namely $\U$. Thus the first $m$ columns of $U_w$,
denoted by $\ket{u_i^w}$, constitute
an orthonormal basis for $\U$, and $MM^*=P_{\U}$, where 
\begin{equation}
\label{eq:iwphi}
P_{\U}=\sum_{i=1}^m \ket{u_i^w} \bra{u_i^w}.
\end{equation}

We now follow the derivation of the previous section, where we
substitute $\Phi_w$ for $\Phi$ and $U_w, V_w$ and $\sigma^w_i$ for $U,
V$ and $\sigma_i$, respectively. The minimizing $\hme_w$ follows 
from Theorem \ref{thm:ls},
\begin{equation}
\label{eq:mw}
\hme_w=\sum_{i=1}^m \ket{u_i^w}\bra{v_i^w}=U_wZ_mV_w^*=
\Phi_w(\Phi_w^*\Phi_w)^{-1/2}=
\Phi W(W^*\Phi^*\Phi W)^{-1/2},
\end{equation}
where the $\ket{v_i^w}$ are the columns of $V_w$.
The resulting error $E'_{min}$ is given by
\begin{equation}
E'_{min}=\sum_{i=1}^m (1-\sigma^w_i)^2.
\end{equation}

Defining $S_w=\Phi_w^*\Phi_w=V_w \Sigma_w^* \Sigma_w V_w^*$, we have
$\tr(S_w)=\sum_{i=1}^m (\sigma_i^w)^2$. In addition,
$S_w=W\Phi^*\Phi W=WSW$. 
Assuming the vectors \phii are normalized, the diagonal elements of
$S$ are all equal to $1$, so $\tr(S_w)=\sum_{i=1}^m w_i^2$ and 
\begin{equation}
E'_{min}=m +\sum_{i=1}^m (w_i^2-2\sigma^w_i).
\end{equation}
From (\ref{eq:esum}) the residual squared error $E_{min}^w$
is therefore given by
\begin{equation}
\label{eq:mwerr}
E^w_{min}=2\sum_{i=1}^m (w_i-\sigma^w_i).
\end{equation}
 
Note that if $W=aI_m$ where $a>0$ is an
arbitrary constant, then $U_w=U$ and $V_w=V$, where $U$ and $V$ are
the unitary matrices in the SVD of $\Phi$. Thus in this case, as we expect,
$\hme_w=\hme$, where \hm is the LSM given by
(\ref{eq:lsmphired}).  

It is interesting to compare the minimal residual squared error
$E_{min}^w$ of (\ref{eq:mwerr})
with the $E_{min}$ of (\ref{eq:emin2}) derived in the previous
section for the non-weighted case, which for the case $r=m$
reduces to $E_{min}=2\sum_{i=1}^m (1-\sigma_i)$.
In the non-weighted case, $w_i=1$ for all $i$, resulting in $W=I$ and
$\tr(W)=m$. Therefore, in order to compare the two cases,
the weights should be
chosen such that $\tr(W)=\sum_{i=1}^m w_i=m$. (Note that only the
ratios of the $w_i$s affect the WLSM. The
normalization $\tr(W)=m$ is chosen for comparison only.) In
this case,  
\begin{equation}
E^w_{min}-E_{min}=2\sum_{i=1}^m (\sigma_i-\sigma^w_i).
\end{equation} 

Recall that $(\sigma_i^w)^2$ and $\sigma_i^{2}$ are the eigenvalues of
$S_w=WSW$ and $S$, respectively.
We may therefore use Ostrowski's theorem (see Appendix A)
to obtain the following bounds:
\begin{equation}
2\bl 1-\max_i w_i\br\sum_{i=1}^m \sigma_i \leq 
E_{min}^w-E_{min} \leq 
2 \bl 1-\min_i w_i \br \sum_{i=1}^m \sigma_i. 
\end{equation} 
Since $\max_i w_i \geq 1$ and $\min_i w_i \leq 1$,
${E}_{min}^w$ can be greater or smaller then $E_{min}$, depending
on the weights $w_i$.

%%%%%%%%%%%%%%%%%%%%%%%%%%%%%%%%%%%%%%%%%%
\section{Example of the LSM and the WLSM}
\label{sec:example}
%%%%%%%%%%%%%%%%%%%%%%%%%%%%%%%%%%%%%%%%%%
We now give an example illustrating the
LSM and the WLSM. \\
Consider the two states,
\begin{equation}
\label{eq:phiex}
\ket{\phi_1}=
\left[
\begin{array}{rr}
1 & 0 
\end{array}
\right]^*, \,\,\,
\ket{\phi_2}=\frac{1}{2}
\left[
\begin{array}{rr}
-1 & \sqrt{3} 
\end{array}
\right]^*.
\end{equation}
We wish to construct the optimal LSM for distinguishing between these
two states. We begin by forming the matrix $\Phi$,
\begin{equation}
\Phi=\frac{1}{2}
\left[
\begin{array}{rr}
2 & -1 \\
0 & \sqrt{3} 
\end{array}
\right].
\end{equation}
The vectors $\ket{\phi_1}$ and $\ket{\phi_2}$ are linearly
independent, so $\Phi$ is a full-rank matrix ($r=2$).
Using Theorem \ref{thm:svd} we may determine the SVD
$\Phi=U\Sigma V^*$, which yields
\begin{equation}
U=\frac{1}{2}
\left[
\begin{array}{rr}
\sqrt{3} & -1 \\
-1 & -\sqrt{3} 
\end{array}
\right],\,\,\,
\Sigma=\frac{1}{\sqrt{2}}
\left[
\begin{array}{rr}
\sqrt{3} & 0 \\
0 & 1 
\end{array}
\right],\,\,\,
V=\frac{1}{\sqrt{2}}
\left[
\begin{array}{rr}
1 & -1 \\
-1 & -1
\end{array}
\right].
\end{equation}
From (\ref{eq:lsmm}) and (\ref{eq:zr}), we now have
\begin{equation}
\hat{M}=UV^*=
\left[
\begin{array}{rr}
0.97 & -0.26 \\
0.26 & 0.97
\end{array}
\right],
\end{equation}
and
\begin{equation}
\label{eq:muex}
\ket{\hat{\mu}_1}=
\left[
\begin{array}{rr}
0.97 & 0.26
\end{array}
\right]^*, \,\,\,
\ket{\hat{\mu}_2}=
\left[
\begin{array}{rr}
-0.26 & 0.97
\end{array}
\right]^*,
\end{equation}
where $\ket{\hat{\mu}_1}$ and $\ket{\hat{\mu}_2}$ 
are the optimal measurement vectors that minimize the
least-squares error defined by (\ref{eq:serror})-(\ref{eq:error}).
Using (\ref{eq:lsmphired}) we may express the optimal
measurement vectors directly in terms of the vectors 
$\ket{\phi_1}$ and $\ket{\phi_2}$,
\begin{equation}
\hat{M}=\Phi(\Phi^*\Phi)^{-1/2}=\Phi
\left[
\begin{array}{rr}
1.12 & 0.30 \\
0.30 & 1.12
\end{array}
\right],
\end{equation}
thus
\begin{equation}
\label{eq:muphiex}
\ket{\hat{\mu}_1}=1.12\ket{\phi_1}+0.30\ket{\phi_2},\,\,\,
\ket{\hat{\mu}_2}=0.30\ket{\phi_1}+1.12\ket{\phi_2}.
\end{equation}

As expected from
Theorem \ref{thm:ls}, $\braket{\hat{\mu}_1}{\hat{\mu}_2}=0$;
the vectors $\ket{\phi_1}$ and $\ket{\phi_2}$
are linearly independent, 
so the optimal measurement vectors must be orthonormal.
The LSM then consists of the orthogonal projection operators
$\Pi_1=\ket{\hat{\mu_1}}\bra{\hat{\mu_1}}$ and
$\Pi_2=\ket{\hat{\mu_2}}\bra{\hat{\mu_2}}$.   

Figure \ref{fig:srmex} depicts the vectors $\ket{\phi_1}$ and
$\ket{\phi_2}$ 
together with the optimal measurement vectors $\ket{\hat{\mu}_1}$ and
$\ket{\hat{\mu}_2}$. 
As is evident from (\ref{eq:muphiex}) and from
Fig.~\ref{fig:srmex}, the optimal measurement vectors are as close as
possible to  
the corresponding states, given that they must be orthogonal.

Suppose now we are given the additional information $p_1=p$ and  $p_2=1-p$,
where $p_1$ and $p_2$ denote the prior probabilities of 
$\ket{\phi_1}$ and $\ket{\phi_2}$ respectively, and  
$p \in (0,1)$. We may still employ the LSM to distinguish
between the two states. However, we expect that a smaller residual squared
error may be achieved by employing a WLSM. 
In Fig.~\ref{fig:wlsmex} we plot the
residual squared error $E_{min}^w$ given by (\ref{eq:mwerr}) as a
function of $p$, when using a  
WLSM with weights  
$w_1=\sqrt{p}$ and $w_2=\sqrt{1-p}$ (we will justify this choice of
weights in \mbox{Section \ref{sec:compare})}.
When $p=1/2$, $w_1=w_2$ and the resulting WLSM 
is equivalent to the LSM. For $p \neq 1/2$, the WLSM does indeed yield a
smaller residual squared error than the LSM (for which the residual
squared error is approximately $0.095$).

\newpage 
%%%%%%%%%%%%%%%%%%%%%%%%%%%%%
\section{Comparison With Other Proposed Measurements}
\label{sec:compare}
%%%%%%%%%%%%%%%%%%%%%%%%%%

We now compare our results with the SRM
proposed by Hausladen \etal in \cite{Haus96}, and with the 
measurement proposed by Peres and Wootters \mbox{in \cite{PW91}}. 

Hausladen \etal construct a POVM  consisting of rank-one 
operators $\Pi_i=\muie \bra{\mu_i}$ to distinguish between
an arbitrary set of vectors \phii. We refer to this POVM as the SRM.
They give two alternative
definitions of their measurement: Explicitly,
\begin{equation}
\label{eq:srmexp}
\overline{M}=((\Phi \Phi^*)^{1/2})^\dagger \Phi,
\end{equation}  
where $\overline{M}$ denotes the matrix of columns $\ket{\overline{\mu}_i}$.
Implicitly, the optimal measurement vectors
$\ket{\overline{\mu_i}}$ are those that satisfy
\begin{equation}
\label{eq:srmimp}
S^{1/2}=\{\braket{\overline{\mu}_j}{\phi_k}\},
\end{equation}   
\ie $\braket{\overline{\mu}_j}{\phi_k}$ is equal to the $jk$th 
element of $S^{1/2}$, where
$S=\Phi^*\Phi$. 

Comparing (\ref{eq:srmexp}) with (\ref{eq:lsmphi2}),
it is evident that the SRM coincides with
the optimal LSM. Furthermore, following the discussion in
Section \ref{sec:CLSM}, if the states are linearly independent
then this measurement is a simple orthogonal measurement and not a more general
POVM. (This observation was made in \cite{SKIH98} as well.)

The implicit definition of (\ref{eq:srmimp}) does not have a unique solution 
when the vectors \phii are linearly dependent. The columns of
$\overline{M}$ are one solution of this equation.
Since the definition depends only on the product 
$M^*\Phi$, any measurement vectors that are columns of $M$ such that
$M^*\Phi=\overline{M}^* \Phi$ constitutes a solution as well. In particular,
the optimal orthogonal LSM $\tilde{M}$ for the linearly dependent case,
given by (\ref{eq:mperp}), satisfies
$\tilde{M}^*\Phi=\overline{M}^* \Phi$, rendering the optimal orthogonal
LSM a solution to (\ref{eq:srmimp}). Consequently, even in the 
case of linearly dependent states, the SRM proposed by Hausladen
\etal and used to achieve the 
classical capacity of a quantum channel may always be chosen as an
orthogonal measurement.   
\mbox{In addition, this measurement is optimal in the least-squares
sense.}

We summarize our results regarding the SRM in the following theorem:
\vspace*{-0.04in}\begin{theorem}[Square-root measurement (SRM)]
\label{thm:srm}
Let $\{ \ket{\phi_i} \}$ be a set of $m$ vectors in an $n$-dimensional complex
Hilbert space $\HH$ that span an $r$-dimensional subspace $\U
\subseteq \HH$. 
Let $\Phi=U \Sigma V^*$ be the rank-$r$ $n \times m$ 
matrix whose columns are the vectors $\ket{\phi_i}$. 
Let $\ket{u_i}$ and $\ket{v_i}$ denote the columns of the unitary matrices $U$
and $V$ respectively, and let $Z_r$ be  defined as in (\ref{eq:zr}).
Let $\{ \ket{\overline{\mu}_i}\}$ be $m$ vectors satisfying  
\[S^{1/2}=\{\braket{\overline{\mu}_j}{\phi_k}\},\]
where $S=\Phi^*\Phi$;
a POVM consisting of the operators 
$\overline{\Pi}_i=\ket{\overline{\mu}_i}\bra{\overline{\mu}_i},\,1
\leq i \leq m$, is referred to as a SRM.
Let $\overline{M}$ be the $n \times m$ 
measurement matrix whose columns are the vectors $\ket{\overline{\mu}_i}$;
$\overline{M}$ is referred to as a SRM matrix. 
Then 
\begin{enumerate}
\item If $r=m$, 
\begin{enumerate}
\item $\overline{M}=\sum_{i=1}^m\ket{u_i}\bra{v_i}=
UZ_mV^*=\Phi(\Phi^*\Phi)^{-1/2}=
((\Phi\Phi^*)^{1/2})^ \dagger\Phi$ is unique; 
\item $\overline{M}^*\overline{M}=I_m$ and the corresponding SRM 
is an orthogonal measurement;
\item the SRM is equal to the optimal LSM.
\end{enumerate}
\item If $r<m$,
\begin{enumerate}
\item the SRM is not unique;
\item $\overline{M}=\sum_{i=1}^m\ket{u_i} \bra{v_i}=UZ_mV^*$ is a SRM
matrix; the corresponding SRM is equal to the optimal orthogonal LSM;
\item define $\overline{M}_\U=P_\U\overline{M}$, where $P_\U$ is a
projection onto $\U$ and $\overline{M}$ is any SRM matrix; then
\begin{enumerate}
\item $\overline{M}_\U$ is unique, and is given by 
$\overline{M}_\U=\sum_{i=1}^r\ket{u_i}\bra{v_i}=
UZ_rV^*=\Phi((\Phi^*\Phi)^{1/2})^{\dagger}=   
((\Phi\Phi^*)^{1/2})^ \dagger\Phi$;
\item $\overline{M}_\U$ is a SRM matrix; the corresponding SRM is equal
to the optimal LSM.
\item $\overline{M}_\U$ may be realized by the optimal orthogonal
LSM \mbox{$\tilde{M}=\sum_{i=1}^m\ket{u_i} \bra{v_i}=UZ_mV^*=\overline{M}$}.
\end{enumerate}
\end{enumerate}
\end{enumerate}
\end{theorem}

The SRM defined in \cite{Haus96} does not take the prior
probabilities of the states \phii into account. In \cite{HW94}, a more
general definition of the SRM that accounts for the  
prior probabilities is given by defining new vectors
$\ket{\phi_i^w}=\sqrt{p_i}\phiie$. The weighted SRM (WSRM) is then
defined as the SRM corresponding to the vectors $\ket{\phi_i^w}$. 
Similarly, the WLSM is equal to the LSM corresponding to the vectors
$w_i\phiie$. Thus, if we  choose the 
the weights $w_i$ proportional to $\sqrt{p_i}$, then the WLSM
coincides with the WSRM.
A theorem similar to Theorem \ref{thm:srm} may then be formulated where 
the WSRM and the WLSM are substituted for the SRM and the LSM.

We next apply our results to a problem
considered by Peres and Wootters in \cite{PW91}. The problem is to
distinguish between three two-qubit states 
\begin{equation}
\label{eq:pwstates}
\ket{\phi_1}=\ket{aa},\,\,
\ket{\phi_2}=\ket{bb},\,\,
\ket{\phi_3}=\ket{cc},
\end{equation}
where $\ket{a}, \ket{b}$ and $\ket{c}$ correspond to
polarizations of a photon at $0^\circ, 60^\circ$ and $120^\circ$, and
the states have equal prior probabilities.
Since the vectors \phii are linearly independent, the
optimal measurement vectors are the columns of \hm given by (\ref{eq:lsmphi}),
\begin{equation}
\label{eq:pwm}
\hme=\Phi(\Phi^*\Phi)^{-1/2}.
\end{equation} 
Substituting (\ref{eq:pwstates})
in (\ref{eq:pwm}) results in the same measurement vectors $\ket{\hat{\mu}_i}$ as
those proposed by Peres and Wootters. Thus their measurement is
optimal in the least-squares sense.
Furthermore, the
measurement that they propose coincides with the SRM for this case. 
In the next section we will show that this measurement also minimizes
the probability of a detection error. 

\newpage
%%%%%%%%%%%%%%%%%%%%%%%%%%%%%
\section{The SRM for Geometrically Uniform State Sets}
\label{sec:minpe}
%%%%%%%%%%%%%%%%%%%%%%%%%%

In this section we will consider the case in which the 
collection of states has a strong symmetry property, called
geometric 
uniformity \cite{F91}.  Under these conditions we show that the SRM is
equivalent to the measurement minimizing the probability of a
detection error, which we refer to as the MPEM.  This result
generalizes a similar result of \mbox{Ban \etal \cite{BKMO97}}.

%%%%%%%%%%%%%%%%%%%%%%%%%%%%%
\subsection{Geometrically Uniform State Sets}
%%%%%%%%%%%%%%%%%%%%%%%%%%

Let $\G$ be a finite abelian (commutative) group of $m$ unitary matrices
$U_i$. That is, $\G$ contains the identity matrix $I$;
if $\G$ contains $U_i$, then it also contains its inverse
$U_i^{-1} = U_i^*$;  the product $U_i U_j$ of any two elements of
$\G$ is in $\G$; and $U_i U_j=U_j U_i$ for any two elements in $\G$
\cite{A88}. 

A state set generated by $\G$ is a set $\SSS = \{\ket{\phi_i} =
U_i\ket{\phi}, U_i \in \G\}$, where
$\ket{\phi}$ is an arbitrary state.  
The group $\G$ will be called the \emph{generating group} of
$\SSS$.
Such a state set has strong
symmetry properties, and will be called
\emph{geometrically uniform} (GU).  For consistency with the symmetry of
$\SSS$, we will assume equiprobable prior probabilities on $\SSS$.

If the group $\G$ contains a rotation $R$ such that $R^k = I$ 
for some integer $k>1$,
then the GU state set $\SSS$ is linearly dependent, because
$\sum_{j=1}^{k} R^j\ket{\phi}$ is a fixed point under
$R$, and the only 
fixed point of a rotation is the zero vector $\ket{0}$.

%A GU state set $\SSS$ is frequently linearly dependent, because the sum
%of all vectors, $(\sum_i U_i)\ket{\phi}$, is a fixed point under
%transformation by any $U_i \in \G$, and therefore is usually the zero
%vector $\ket{0}$.

Since $U_i^* = U_i^{-1}$, the inner product of two  vectors in $\SSS$ is
\begin{equation}
\braket{\phi_i}{\phi_j} = \bra{\phi}U_i^{-1} U_j\ket{\phi} =
s(U_i^{-1}U_j),
\end{equation}
where $s$ is the function on $\G$ defined by
\begin{equation}
s(U_i) = \bra{\phi}U_i\ket{\phi}.
\end{equation}
For fixed $i$, the set $U_i^{-1}\G = \{U_i^{-1}U_j, U_j \in \G\}$ 
is just a permutation of $\G$ since $U_i^{-1}U_j \in \G$
for all $i,j$ \cite{A88}.  Therefore
the $m$ numbers $\{s(U_i^{-1}U_j), 1 \leq j \leq m\}$ are a permutation
of the  numbers $\{s(U_i), 1 \leq i \leq m\}$. The same is true for fixed
$j$. Consequently, every row and column of 
the $m \times m$ Gram matrix $S =
\{\braket{\phi_i}{\phi_j}\}$ 
is a permutation of the numbers $\{s(U_i), 1 \le i \le m\}$. 

It will be convenient to replace the multiplicative group $\G$ by an
additive group $G$ to which $\G$ is isomorphic\footnote{
Two groups $\G$ and $\G'$ are {\em isomorphic}, denoted by $\G \cong
\G'$, if there is a bijection (one-to-one and onto map)
$\varphi: \G \to \G'$ which satisfies $\varphi(xy)=\varphi(x)\varphi(y)$
for all $x,y \in \G$ \cite{A88}.}.
Every finite abelian group $\G$ is isomorphic
to a direct product $G$
of a finite number of cyclic groups:  $\G \cong G =
\Z_{m_1} \times \cdots \times \Z_{m_p}$, where $\Z_{m_k}$ is the cyclic
additive group of integers modulo $m_k$, and $m = \prod_k m_k$
\cite{A88}.   Thus every element $U_i \in \G$ can be associated with an
element $g \in G$ of the form $g = (g_1,g_2,\ldots ,g_p)$, where $g_k \in
\Z_{m_k}$. We denote this one-to-one correspondence by $U_i \lra g$.
Because the correspondence is an isomorphism, it follows that if $U_i
\lra g,\,U_k \lra g',\,U_l \lra g''$ and $U_i=U_kU_l$, then $g=g' + g''$,
where the addition of $g' = (g'_1,g'_2,\ldots ,g'_p)$ and $g'' =
(g''_1,g''_2,\ldots ,g''_p)$ is performed by componentwise addition
modulo the corresponding $m_k$.

Each state vector
$\ket{\phi_i}=U_i\ket{\phi}$ will henceforth be denoted as
$\ket{\phi(g)}$, where 
$g \in G$ is the group element corresponding to $U_i \in \G$.
The zero element $0=(0,0,\ldots,0) \in G$
corresponds to the identity matrix $I \in \G$, and an additive inverse
$-g \in G$ corresponds to a multiplicative inverse $U_i^{-1} =
U_i^* \in \G$.
The Gram matrix is then the $m \times m$ matrix 
\begin{equation}
\label{eq:sij}
S = \{\braket{\phi(g')}{\phi(g)}, g', g \in G\} = \{s(g - g'),
g', g \in G\},
\end{equation}
with row and column indices $g', g \in G$, where $s$ is now the function
on $G$ defined by
\begin{equation}
s(g) = \braket{\phi(0)}{\phi(g)}.
\end{equation}

%%%%%%%%%%%%%%%%%%%%%%%%%%%%%
\subsection{The SRM}
%%%%%%%%%%%%%%%%%%%%%%%%%%

We now obtain the SRM for a GU state set. We begin by determining the
SVD of $\Phi$.
To this end we introduce the following definition.
The Fourier transform (FT) of a complex-valued function $\varphi: G \to
\C$ defined on $G = \Z_{m_1} \times \cdots
\times
\Z_{m_p}$ is the complex-valued function $\hat{\varphi}: G \to \C$
defined by
\begin{equation}
\label{eq:fh}
\hat{\varphi}(h) = \frac{1}{\sqrt{m}}\sum_{g \in G} \inner{h}{g} \varphi(g),
\end{equation}
where the Fourier kernel $\inner{h}{g}$ is
\begin{equation}
\label{eq:gh}
\inner{h}{g} = \prod_{k=1}^p e^{-2 \pi i h_kg_k/m_k}.
\end{equation}
Here $h_k$ and $g_k$ are the $k$th components of $h$ and $g$
respectively, and the product $h_kg_k$ is taken as an ordinary integer
modulo $m_k$.  The Fourier kernel evidently satisfies:
\begin{eqnarray}
\label{eq:ghprop}
\inner{h}{g} & = & \inner{g}{h}; \\
\inner{h}{g}^* & = & \inner{-h}{g} = \inner{h}{-g}; \\
\label{eq:sghprop}
\inner{h + h'}{g} & = & \inner{h}{g}\inner{h'}{g}; \\
\label{eq:shgprop}
\inner{h}{g + g'} & = & \inner{h}{g}\inner{h}{g'}.
\end{eqnarray}

We define the FT matrix over $G$ as the $m
\times m$ matrix $\FF = \{\frac{1}{\sqrt{m}}\inner{h}{g}, h,g \in G\}$.  The FT
of a column vector $\ket{\varphi} = \{\varphi(g), g \in G\}$ is then the
column vector $\ket{\hat{\varphi}} = \{\hat{\varphi}(h), h \in G\}$
given by $\ket{\hat{\varphi}} = \FF\ket{\varphi}$.
It is easy to show that the rows and columns of $\FF$ are orthonormal;
\ie $\FF$ is unitary:
\begin{equation}
\FF^*\FF = \FF\FF^* = I_m.
\end{equation}
Consequently we obtain the inverse FT formula
\begin{equation}
\ket{\varphi} = \FF^*\ket{\hat{\varphi}} =
\left\{\frac{1}{\sqrt{m}}\sum_{h \in G} 
\inner{h}{g}^* \hat{\varphi}(h),g \in G\right\}.
\end{equation}

We now show that the eigenvectors of the Gram matrix $S$
of (\ref{eq:sij}) are the column vectors $\ket{\FF(h)} =
\{\frac{1}{\sqrt{m}}\inner{h}{g}, g \in G\}$ of $\FF$.   Let $\bra{S(g')} = \{s(g - g'), g
\in G\}$ be the $g'$th row of $S$. Then 
\begin{equation}
\braket{S(g')}{\FF(h)} = \frac{1}{\sqrt{m}}\sum_{g \in G} \inner{h}{g}
s(g - g')  = 
\frac{1}{\sqrt{m}}\sum_{g'' \in G} \inner{h}{g' + g''} s(g'')
=\inner{h}{g'}\hat{s}(h), 
\end{equation}
where the last equality follows from (\ref{eq:shgprop}), and
$\{\hat{s}(h), h \in G\}$ is the FT of $\{s(g), g \in G\}$.
Thus $S$ has the eigendecomposition
\begin{equation}
S = \FF \overline{\Sigma}^2 \FF^*,
\end{equation}
where $\overline{\Sigma}$ is an $m \times m$ diagonal matrix with diagonal
elements $\{\sigma(h) = m^{1/4}\sqrt{\hat{s}(h)}, h \in G\}$ (the eigenvalues
$\sigma^2(h)$ are real and nonnegative because $S$ is
Hermitian).
Consequently, the $V$-basis of the SVD of $\Phi$ is $V = \FF$, and
the singular values of $\Phi$ are $\sigma(h)$. 

We now write the SVD of $\Phi$ in the following form:
\begin{equation}
\label{eq:svdalt}
\Phi = \Upsilon \overline{\Sigma} \FF^* = \sum_{h \in G}
\sigma(h)\ket{u(h)}\bra{\FF^*(h)},
\end{equation}
where
$\Upsilon$ is the $n \times m$ matrix whose columns $\ket{u(h)}$ are
the columns of the $U$-basis of the SVD of $\Phi$ for values of $h \in G$
such that $\sigma(h) \neq 0$ and are zero columns otherwise,
and $\FF^* = \{\frac{1}{\sqrt{m}}\inner{h}{g}^*, h, g \in G\}$ has
rows $\bra{\FF^*(h)} 
= \{\frac{1}{\sqrt{m}}\inner{h}{g}^*, g \in G\}$.
It then follows that
\begin{eqnarray}
\label{eq:ubasis}
\ket{u(h)} & = & \left\{\begin{array}{ll}
\Phi \ket{\FF(h)}/\sigma(h) = \ket{\hat{\phi}(h)}/\sigma(h),  &
 \mbox{if}\,\, \sigma(h) \neq 0; \\
\ket{0}, & \mathrm{otherwise,}
\end{array}\right.
\end{eqnarray}
where
\begin{equation}
\label{eq:ftphi}
\ket{\hat{\phi}(h)} = \frac{1}{\sqrt{m}}\sum_{g \in G}
\inner{h}{g}\ket{\phi(g)} 
\end{equation}
is the $h$th element of the FT of $\Phi$ regarded as a
row vector of column vectors, \mbox{$\Phi = \{\ket{\phi(g)}, g \in G\}$}.

Finally, the SRM is given by the measurement matrix
\begin{equation}
\label{eq:mgu}
M = \Upsilon \FF^* = \sum_{h \in G}\ket{u(h)}\bra{\FF^*(h)}.
\end{equation}
The measurement vectors $\ket{\mu(g)}$
(the columns of $M$) are thus the inverse
FT of the columns of $\Upsilon$:
\begin{equation}
\label{eq:mu}
\ket{\mu(g)} = \frac{1}{\sqrt{m}}\sum_{h \in G}\inner{g}{h}^*\ket{u(h)}.
\end{equation}

Note that if $\ket{\phi(g)} = U_i\ket{\phi}$ where $U_i \lra g$, and
$U_j \lra g'$, then  $U_j\ket{\phi(g)} = U_jU_i\ket{\phi} = \ket{\phi(g +
g')}$.  Therefore left multiplication of the state vectors $\Phi =
\{\ket{\phi(g)}, g \in G\}$ by $U_j$ permutes the state vectors to
$U_j\Phi = \{\ket{\phi(g + g')}, g \in G\}$.  We now show that under
this transformation the measurement vectors are similarly permuted;  \ie
\mbox{$U_j M = \{\ket{\mu(g + g')}, g \in G\}$}.  The FT of
the permuted vectors $\{\ket{\phi(g + g')}, g \in G\}$ is
\begin{equation}
\label{eq:perm1}
\ket{\hat{\phi'}(h)} = \frac{1}{\sqrt{m}}\sum_{g \in G}
\inner{h}{g}\ket{\phi(g + g')} = 
\frac{1}{\sqrt{m}} \sum_{g'' \in G} \inner{h}{g'' - g'}\ket{\phi(g'')}  =
\inner{h}{g'}^* \ket{\hat{\phi}(h)}.
\end{equation}
Normalization by $\sigma(h)^{-1}$ when $\sigma(h)
\neq 0$ yields $\ket{u'(h)} = \inner{h}{g'}^* \ket{u(h)}$.  Finally, the
inverse FT yields the measurement vectors
\begin{equation}
\label{eq:perm2}
\ket{\mu'(g)} = \frac{1}{\sqrt{m}}\sum_{h \in G}\inner{g}{h}^*\ket{u'(h)}
= \frac{1}{\sqrt{m}}\sum_{h \in G}\inner{g + g'}{h}^*\ket{u(h)} =
\ket{\mu(g + g')}, 
\end{equation}
where we have used (\ref{eq:ghprop}) and (\ref{eq:sghprop}). 

This shows that the measurement vectors $\ket{\mu(g)}$ have the same
symmetries as the state vectors;  \ie they also form a GU set with
generating group $\G$.  Explicitly, if $U_i \lra g$, then $\ket{\mu(g)} =
U_i\ket{\mu}$, where  $\ket{\mu}$ denotes $\ket{\mu(0)}$.

%%%%%%%%%%%%%%%%%%%%%%%%%%%%%
\subsection{The SRM and the MPEM}
%%%%%%%%%%%%%%%%%%%%%%%%%%

We now show that for GU state sets the SRM is equivalent to the MPEM.
In the process, we derive a sufficient condition for the SRM to
minimize the probability of a detection error for a general state set
(not necessarily GU) comprised of linearly independent states.

Holevo \cite{Hol73,Helb} and Yuen \etal \cite{Ken75} showed that a set of measurement operators
$\Pi_i$ comprises the 
MPEM for a set of weighted
density operators $W_i = p_i\rho_i$ if they
satisfy 
\begin{eqnarray}
\label{eq:holevo}
\Pi_i(W_j - W_i)\Pi_j &=& \zerob ,\,\, \forall g,g'; \\
\label{eq:holevopd}
\Gamma- W_i &\geq& \zerob ,\,\, \forall g,
\end{eqnarray}
where 
\begin{equation}
\label{eq:gamma}
\Gamma=\sum_{j=1}^m \Pi_j W_j
\end{equation}
and is required to be Hermitian. Note that if (\ref{eq:holevo}) is satisfied,
then $\Gamma$ is Hermitian. 

In our case the measurement operators $\Pi_i$ are the operators
$\ket{\mu(g)}\bra{\mu(g)}$, and the  weighted density
operators may be taken simply as the projectors
$\ket{\phi(g)}\bra{\phi(g)}$, since their prior probabilities are equal.
The conditions (\ref{eq:holevo})-(\ref{eq:holevopd}) then become
\begin{eqnarray}
\label{eq:holevo2}
&& \ket{\mu(g)}\braket{\mu(g)}{\phi(g')}\braket{\phi(g')}{\mu(g')}
\bra{\mu(g')} =
\ket{\mu(g)}\braket{\mu(g)}{\phi(g)}\braket{\phi(g)}{\mu(g')}
\bra{\mu(g')},\,\, \forall g,g'; \\
\label{eq:holevopd2}
&& \sum_{g'} \ket{\mu(g')}\braket{\mu(g')}{\phi(g')}\bra{\phi(g')}-
\ket{\phi(g)}\bra{\phi(g)} \geq \zerob ,\,\, \forall g.
\end{eqnarray}

We first verify that the conditions (\ref{eq:holevo}) (or equivalently
(\ref{eq:holevo2})) are satisfied.
Since the matrix $M^* \Phi  =
\FF \overline{\Sigma} \FF^*$ is symmetric,
$\braket{\mu(g')}{\phi(g)} = \bra{\mu}U_j^{-1}U_i\ket{\phi} = w(g - g')$,
where $w(g) = \braket{\mu}{\phi(g)}$ is a complex-valued function that
satisfies $w(-g) = w^*(g)$.  Therefore,
\begin{eqnarray}
\braket{\mu(g)}{\phi(g')} & = & w(g' - g)= w^*(g -g')=
\braket{\phi(g)}{\mu(g')}; \\ 
\braket{\phi(g')}{\mu(g')} & = & w^*(0) = w(0)=\braket{\mu(g)}{\phi(g)}.
\end{eqnarray}
Substituting these relations back into (\ref{eq:holevo2}), we obtain
\begin{equation}
w(0)w(g' - g)\ket{\mu(g)}\bra{\mu(g')} =
w(0)w(g' - g)\ket{\mu(g)}\bra{\mu(g')} ,\,\, \forall g,g',
\end{equation}
which verifies that the conditions (\ref{eq:holevo}) are satisfied.  

Next, we show that conditions (\ref{eq:holevopd}) are satisfied. Since
$M^* \Phi=\FF\overline{\Sigma} \FF^*$, 
\begin{equation}
\label{eq:c}
w(0)=\braket{\mu(g)}{\phi(g)}=
\bra{\FF(g)}\overline{\Sigma}\ket{\FF(g)},
\end{equation}
where $\bra{\FF(g)}$ denotes the row of $\FF$ corresponding to $g$.
Then,
\begin{equation}
\label{eq:ngamma}
\Gamma=\sum_{g'} \ket{\mu(g')}\braket{\mu(g')}{\phi(g')}\bra{\phi(g')}
= w(0)\sum_{g'} \ket{\mu(g')} \bra{\phi(g')}.
\end{equation}
From (\ref{eq:svdalt}) and (\ref{eq:mgu}) we have
\begin{equation}
\sum_{g'} \ket{\mu(g')} \bra{\phi(g')}=\Upsilon \overline{\Sigma} \Upsilon^*,
\end{equation}
and
\begin{equation}
\label{eq:outerphi}
\ket{\phi(g)} \bra{\phi(g)}=\Upsilon \overline{\Sigma}
\ket{\FF(g)}\bra{\FF(g)} 
\overline{\Sigma}\Upsilon^*.
\end{equation}

Substituting (\ref{eq:ngamma})-(\ref{eq:outerphi}) 
back into  (\ref{eq:holevopd2}), the conditions of (\ref{eq:holevopd2})
reduce to
\begin{equation}
\Upsilon \bl w(0) \overline{\Sigma} 
-  \overline{\Sigma}\ket{\FF(g)}\bra{\FF(g)}
\overline{\Sigma} \br \Upsilon^*  \geq \zerob,
\end{equation}
where $w(0)$ is given by (\ref{eq:c}).
It is therefore sufficient to show that 
\begin{equation}
T=w(0)\overline{\Sigma}-\overline{\Sigma}\ket{\FF(g)}\bra{\FF(g)}
\overline{\Sigma} \geq \zerob 
\end{equation}
or equivalently that $\braket{u|T}{u} \geq 0$ for any $\ket{u} \in \C^m$.
Using the Cauchy-Schwartz inequality we have
\begin{eqnarray}
\braket{u|T}{u}
& = & \bra{\FF(g)}\overline{\Sigma}\ket{\FF(g)}
\bra{u}\overline{\Sigma} \ket{u} -
\bra{u}\overline{\Sigma}\ket{\FF(g)}\bra{\FF(g)}
\overline{\Sigma}\ket{u} \nonumber \\
& \geq & \bra{\FF(g)}\overline{\Sigma}\ket{\FF(g)}
\bra{u}\overline{\Sigma}\ket{u}-
\bra{\FF(g)}\overline{\Sigma}\ket{\FF(g)} 
\bra{u}\overline{\Sigma} \ket{u} 
=0,
\end{eqnarray}
which verifies that the conditions (\ref{eq:holevopd}) are satisfied.  
We conclude
that when the state set $\SSS$ is GU, the SRM is also the MPEM.

An alternative way of deriving this result for the case of linearly
independent states $\ket{\phi_i}$ is by use of the following
criterion of \mbox{Sasaki \etal \cite{SKIH98}}.
Denote by $\Phi_w$ the matrix whose columns are the vectors 
$\ket{\phi_i^w}=\sqrt{p_i}\ket{\phi_i}$ where $p_i$ is the prior
probability of state 
$i$. If the states are linearly independent and 
\mbox{$S^{1/2}=(\Phi_w^*\Phi_w)^{1/2}$} has constant
diagonal elements, 
then the SRM corresponding to the vectors $\ket{\phi_i^w}$ (\ie a WSRM),
is equivalent to the MPEM.

This condition is hard to verify directly from the
vectors $\ket{\phi_i^w}$.
The difficulty arises from the fact
that generally there is no simple relation between 
the diagonal elements of $S^{1/2}$ and the elements of
$S$. Thus given an 
ensemble of pure states \phii with prior probabilities $p_i$, 
we typically need to calculate $S^{1/2}$ (which in itself is not
simple to do analytically) in order to verify the condition above. 
However, as we now show, 
in some cases this condition may  be
verified directly from the elements of $S$ using the SVD.

Employing the SVD $\Phi_w=U\Sigma V^*$ we may express $S^{1/2}$ as
\begin{equation}
S^{1/2}=(\Phi_w^*\Phi_w)^{1/2}=V(\Sigma^*\Sigma)^{1/2} V^*=
V\overline{\Sigma}V^*,
\end{equation}
where $\overline{\Sigma}$ is a diagonal matrix with the first $r$
diagonal elements equal to $\sigma_i$, and the remaining elements
all equal zero, where the $\sigma_i$ are the singular values of $\Phi_w$.
Thus, the WSRM is equal to the MPEM if $\bra{\overline{v}_i}
\overline{\Sigma}\ket{\overline{v}_i}=c,\,\,1 \leq i \leq m$, where the vectors
$\ket{\overline{v}_i}$ denote the columns of 
$V^*$, and $c$ is a constant.
In particular, if the elements of $V$ all have equal magnitude,
then  $\bra{\overline{v}_i} \overline{\Sigma}\ket{\overline{v}_i}$ is constant,
and the SRM minimizes the probability of a detection error.

If the state set $\SSS$ is GU,
then the matrix $V$ is the FT matrix $\FF$, whose elements
all have magnitude equal to one. Thus, if the states are
linearly independent and GU, then 
the SRM is equivalent to the MPEM.

We summarize our results regarding GU state sets in the following theorem:
\begin{theorem}[SRM for GU state sets]
\label{thm:gu}
Let $\SSS = \{\ket{\phi_i} =
U_i\ket{\phi}, U_i \in \G\}$, 
be a geometrically uniform state set generated
by a finite abelian group $\G$ of unitary matrices, where
$\ket{\phi}$ is an arbitrary state. Let $\G \cong
G$, and let  $\Phi$ be
the matrix of columns $\ket{\phi_i}$. Then the SRM is given by the measurement
matrix
\[
M = \Phi \FF \overline{\Sigma}^\dagger \FF^* = \sum_{h \in
G}\ket{u(h)}\bra{\FF^*(h)},
\]
where $\FF$ is the Fourier transform matrix over $G$,
$\overline{\Sigma}^\dagger$ is the diagonal matrix whose diagonal elements are
$\sigma(h)^{-1}$ when
$\sigma(h) \neq 0$ and $0$ otherwise, where $\{\sigma(h), h \in G\}$ are
the singular values of $\Phi$,
$\ket{u(h)} = \ket{\hat{\phi}(h)}/\sigma(h)$ when
$\sigma(h) \neq 0$ and $\ket{0}$ otherwise, where $\{\ket{\hat{\phi}(h)}, h \in
G\}$ is the Fourier transform of $\{\ket{\phi(g)}, g \in G\}$, and
$\bra{\FF^*(h)}$ is the $h$th row of $\FF^*$.

The SRM has the following
properties:
\begin{enumerate}
\item \label{thm:gusym}
The measurement matrix $M$ has the same symmetries as $\Phi$;
\item The SRM is the least-squares measurement (LSM);
\item The SRM is the minimum-probability-of-error measurement (MPEM).
\end{enumerate}
\end{theorem}

%%%%%%%%%%%%%%%%%%%%%%%%%%%%%%%%%%%%%%%%
\subsection{Example of a GU State Set}
%%%%%%%%%%%%%%%%%%%%%%%%%%%%%%%%%%%%%%
We now consider an example demonstrating the ideas of the previous
section. Consider the group $\G$ of $m=4$ unitary matrices $U_i$,
where
\begin{equation}
U_1=I_4,\,\,\,
U_2=\left[
\begin{array}{rrrr}
-1 & 0 & 0 & 0 \\
0 & 1 & 0 & 0 \\
0 & 0 & -1 & 0 \\
0 & 0 & 0 & -1
\end{array}
\right],\,\,\,
U_3=\left[
\begin{array}{rrrr}
-1 & 0 & 0 & 0 \\
0 & -1 & 0 & 0 \\
0 & 0 & 1 & 0 \\
0 & 0 & 0 & -1
\end{array}
\right],\,\,\,
U_4=U_2U_3.
\end{equation}
Let the state set be $\SSS=\{\ket{\phi_i}=U_i\ket{\phi},\,\,1 \leq i \leq
4\}$,  where
$\ket{\phi}=\frac{1}{2}[1\,\, 1\,\, 1\,\, 1]^*$. Then $\Phi$ is
\begin{equation}
\label{eq:phigu}
\Phi=\frac{1}{2}\left[
\begin{array}{rrrr}
1 & -1 & -1 & 1 \\
1 & 1 & -1 & -1 \\
1 & -1 & 1 & -1 \\
1 & -1 & -1 & 1
\end{array}
\right],
\end{equation}
and the Gram matrix $S$ is given by
\begin{equation}
\label{eq:sgu}
S=\frac{1}{2}
\left[
\begin{array}{rrrr}
2 & -1 & -1 & 0 \\
-1 & 2 & 0 & -1 \\
-1 & 0 & 2 & -1 \\
0 & -1 &  -1 & 2
\end{array}
\right].
\end{equation}
Note that the sum of the states $\ket{\phi_i}$ is $\ket{0}$, so the
state set is linearly dependent.

In this case $\G$ is isomorphic to $G=\Z_2 \times \Z_2$, \ie
$G=\{(0,0),(0,1),(1,0),(1,1)\}$.
The multiplication table of the group $\G$ is
\begin{equation}
\label{eq:utable}
\begin{array}{c|cccc}
 & U_1 & U_2 & U_3 & U_4 \\
\hline
U_1 & U_1 & U_2 & U_3 & U_4 \\
U_2 & U_2 & U_1 & U_4 & U_3 \\
U_3 & U_3 & U_4 & U_1 & U_2 \\
U_4 & U_4 & U_3 & U_2 & U_1.
\end{array}
\end{equation}
If we define the correspondence
\begin{equation}
\label{eq:coress}
U_1 \lra (0,0),\,\,U_2 \lra (0,1),\,\, U_3 \lra (1,0),\,\,U_4 \lra (1,1),
\end{equation}
then this table becomes the
addition table of $G=\Z_2 \times \Z_2$:
\begin{equation}
\label{eq:atable}
\begin{array}{c|cccc}
 & (0,0) & (0,1) & (1,0) & (1,1) \\
\hline
(0,0) & (0,0) & (0,1) & (1,0) & (1,1) \\
(0,1) & (0,1) & (0,0) & (1,1) & (1,0) \\
(1,0) & (1,0) & (1,1) & (0,0) & (0,1) \\
(1,1) & (1,1) & (1,0) & (0,1) & (0,0).
\end{array}
\end{equation}
Only the way in which the elements are labeled distinguishes the table of
(\ref{eq:atable}) from the table of (\ref{eq:utable});
thus $\G \cong G$.
Comparing (\ref{eq:utable}) and (\ref{eq:atable})
with (\ref{eq:sgu}), we see that the tables and the matrix $S$ have the
same symmetries.

Over $G=\Z_2 \times \Z_2$, the Fourier matrix $\FF$ is the
Hadamard matrix
\begin{equation}
\FF=\frac{1}{2}\left[
\begin{array}{rrrr}
1 & 1 & 1 & 1 \\
1 & -1 & 1 & -1 \\
1 & 1 & -1 & -1 \\
1 & -1 & -1 & 1
\end{array}
\right].
\end{equation}

Using (\ref{eq:ubasis}) and
(\ref{eq:mgu}), we may find the measurement matrix of the SRM:
\begin{equation}
\label{eq:mgun}
M =\frac{1}{2\sqrt{2}}\left[
\begin{array}{rrrr}
1 & -1 & -1 & 1 \\
\sqrt{2} & \sqrt{2} & -\sqrt{2} & -\sqrt{2} \\
\sqrt{2} & -\sqrt{2} & \sqrt{2} & -\sqrt{2} \\
1 & -1 & -1 & 1
\end{array}
\right].
\end{equation}
We verify that the columns $\ket{\mu_i}$ of $M$ may be
expressed as $\ket{\mu_i}=U_i\ket{\mu_1},\,\,1 \leq i \leq 4$, where
$\ket{\mu_1}=\frac{1}{2\sqrt{2}}[1\,\, \sqrt{2}\,\, \sqrt{2}\,\,
1]^*$.  Thus the 
measurement vectors $\ket{\mu_i}$ also form a GU set generated by $\G$.

%%%%%%%%%%%%%%%%%%%%%%%%%%%%%%%%%%%%
\subsection{Applications of GU State Sets}
%%%%%%%%%%%%%%%%%%%%%%%%%%%%%%%%%%%%
We now discuss some applications of Theorem
\ref{thm:gu}.

{\em A. Binary state set:} 
Any binary state set $\SSS=\{\ket{\phi_1},\ket{\phi_2}\}$ is GU, because it can
be generated by the binary group $\G = \{I, R\}$, where $I$ is the
identity and $R$ is the reflection about the hyperplane halfway between
the two states. 
Specifically, if the two states $\ket{\phi_1}$ and
$\ket{\phi_2}$ are real, then
\begin{equation}
\label{eq:R}
R = I - 2\frac{\ket{w}\bra{w}}{\braket{w}{w}},
\end{equation}
where $\ket{w}=\ket{\phi_2}-\ket{\phi_1}$.
We may immediately verify that $R^2=I$, so that $R^{-1}=R$, and that
$\ket{\phi_2}=R\ket{\phi_1}$. 

If the states are complex with
$\braket{\phi_1}{\phi_2}=ae^{j\theta}$, then define 
$\ket{\phi_2'}=e^{-j\theta}\ket{\phi_2}$. The states $\ket{\phi_2}$ and
$\ket{\phi_2'}$ differ by a phase factor and therefore
correspond to the same  physical state. We may therefore replace our
state set $\SSS = \{\ket{\phi_1},\ket{\phi_2}\}$ by the equivalent
state set $\SSS = \{\ket{\phi_1},\ket{\phi_2'}\}$. Now the generating
group is $\G = \{I, R\}$, where $R$ is defined by (\ref{eq:R}), with
$\ket{w}=\ket{\phi_2'}-\ket{\phi_1}$. 

The generating group $\G = \{I, R\}$ is isomorphic to $G=\Z_2$. The
Fourier matrix $\FF$ therefore reduces to the $2 \times 2$ discrete FT
(DFT) matrix,
\begin{equation}
\label{eq:fbin}
\FF=\frac{1}{\sqrt{2}}\left[
\begin{array}{rr}
1 & 1 \\
1 & -1 
\end{array}
\right].
\end{equation}
The squares of the singular values of $\Phi$ are therefore
$\{\sigma^2(h)=\sqrt{2}\hat{s}(h),h \in G\}$ where  $\{\hat{s}(h),h \in G\}$
are the DFT values of $\{s(g),g\in G\}$, with $s(0)=1$ and $s(1)=a$.
Thus, 
\begin{eqnarray}
\label{eq:sbin}
\sigma^2(0)=1+a; \nonumber \\
\sigma^2(1)=1-a.
\end{eqnarray}
From Theorem \ref{thm:gu} we then have
\begin{equation}
\label{eq:gubin}
M = \Phi \FF \overline{\Sigma}^\dagger \FF^* =\frac{1}{2} \Phi 
\left[
\begin{array}{rr}
\frac{1}{\sigma(0)}+\frac{1}{\sigma(1)} &
\frac{1}{\sigma(0)}-\frac{1}{\sigma(1)}\\ 
\frac{1}{\sigma(0)}-\frac{1}{\sigma(1)} &
\frac{1}{\sigma(0)}+\frac{1}{\sigma(1)} 
\end{array}
\right].
\end{equation}

We may now apply (\ref{eq:gubin}) to the example of Section
\ref{sec:example}. In that example 
$a=\braket{\phi_1}{\phi_2}=-1/2$.  
From (\ref{eq:sbin}) it then follows that $\sigma(0)=1/\sqrt{2}$ and
$\sigma(1)=\sqrt{3/2}$. Substituting these values in  
(\ref{eq:gubin}) yields
\begin{equation}
M=\Phi 
\left[
\begin{array}{rr}
1.12 & 0.30 \\
0.30 & 1.12
\end{array}
\right],
\end{equation}
which is equivalent to the optimal measurement matrix obtained in Section
\ref{sec:example}. 

We could have obtained the measurement vectors directly from the
symmetry property 
of \mbox{Theorem \ref{thm:gu}.\ref{thm:gusym}}. The state set
$\SSS=\{\ket{\phi_1},\ket{\phi_2}\}$ is
invariant under 
a reflection about the line halfway between
the two states, as illustrated in Fig.~\ref{fig:guex}. The measurement
vectors must also be invariant under the same reflection. In addition,
since the states are linearly independent, the measurement vectors 
must be orthonormal. This completely determines the measurement vectors
shown in Fig.~\ref{fig:guex}. (The only other possibility, namely the 
negatives of these two vectors, is physically equivalent.)

{\em B. Cyclic state set:} 
A cyclic generating group $\G$ has elements $U_i = Q^{i-1},
1 \le i \le m$, where $Q$ is a unitary matrix with $Q^m = I$.  A cyclic
group generates a cyclic state set $\SSS =
\{\ket{\phi_i}=Q^{i-1}\ket{\phi},\,\,1
\leq i \leq m\}$, where $\ket{\phi}$ is arbitrary.  Ban \etal
\cite{BKMO97} refer to such a cyclic state set as a symmetrical state set,
and show that in that case the SRM is equivalent to the MPEM.
This result is a special case of Theorem \ref{thm:gu}.

Using Theorem \ref{thm:gu}  we may obtain the measurement matrix $M$
as follows.
If $\G$ is cyclic, then $S$ is a circulant
matrix\footnote{A circulant matrix is a  matrix where every row (or
column) is obtained by a right circular shift (by one position) of the
previous row (or column). An example is:
$\left[ \begin{array}{ccc}
a_0 & a_2 & a_1 \\
a_1 & a_0 & a_2 \\
a_2 & a_1 & a_0
\end{array} \right].$}, and
$G$ is the cyclic group $\Z_m$. The FT kernel is then
$\inner{h}{g} = e^{-2 \pi ihg/m}$ for $h,g \in \Z_m$, and
the Fourier matrix $\FF$ reduces to the $m \times m$ DFT matrix.
The singular values of $\Phi$ are $m^{1/4}$ times the square roots of the DFT
values of the inner products $\{\braket{\phi_1}{\phi_j},1 \leq j \leq m\}$.
We then calculate $M=\Phi \FF \overline{\Sigma}^\dagger \FF^*$.

{\em C. Peres-Wootters measurement:} We may apply these results to the
Peres-Wootters problem considered 
at the end of Section \ref{sec:compare}.  
In this problem the states to be distinguished are given by 
$\ket{\phi_1}=\ket{aa}, \ket{\phi_2}=\ket{bb}$ and
$\ket{\phi_3}=\ket{cc}$, 
where $\ket{a}, \ket{b}$ and $\ket{c}$ correspond to
polarizations of a photon at $0^\circ, 60^\circ$ and $120^\circ$, and
the states have equal prior probabilities.
The state set $\SSS=\{\ket{\phi_1},\ket{\phi_2},\ket{\phi_3}\}$ is
thus a cyclic state set with $\ket{\phi_i}=U_i\ket{\phi_1},1 \leq i
\leq 3$, where $U_i=(Q \otimes Q)^{i-1}$ and $Q$ is a rotation by 
$60^\circ$.

In Section \ref{sec:compare} we concluded that the Peres-Wootters
measurement is equivalent to the SRM and consequently minimizes the
squared error.
From Theorem \ref{thm:gu} we now conclude that the
Peres-Wootters measurement 
minimizes the probability of a detection error as well. 

%%%%%%%%%%%%%%%%%%%%%%%%%%%%%%%%%%%%
\section{Conclusion}
%%%%%%%%%%%%%%%%%%%%%%%%%%%%%%%%%%%%
In this paper we constructed optimal measurements in the least-squares
sense for distinguishing
between a collection of quantum states.
We considered POVMs consisting of rank-one operators, where the
vectors were chosen to
minimize a  possibly weighted sum 
of squared errors. We saw that for linearly independent states the
optimal least-squares measurement is an orthogonal 
measurement, which coincides with the 
SRM proposed by Hausladen \etal \cite{Haus96}. 
If the states are linearly dependent, then the optimal POVM still has
the same general form. 
We showed that it may be realized
by an orthogonal measurement of the same form as in the linearly
independent case.   
We also noted that the SRM, which was constructed by Hausladen \etal
\cite{Haus96} and used to
achieve the classical channel capacity of a quantum channel, may always
be chosen as an orthogonal measurement. 

We showed that for a GU state set the SRM minimizes the probability
of a detection error. We also derived
a sufficient condition for the SRM to minimize the
probability of a detection error in the case of linearly independent
states based on the properties of the SVD.

%%%%%%%%%%%%%%%%%%%%%%%%%%%%%%%%
\section*{Acknowledgments}
%%%%%%%%%%%%%%%%%%%%%%%%%%%%%%%%

We are grateful to A. S. Holevo and H. P. Yuen for helpful comments.  The
first author wishes to thank A. V. Oppenheim for his encouragement and
support.

\newpage
%%%%%%%%%%%%%%%%%%%%%%%%%%%%%%%%
\appendix
%%%%%%%%%%%%%%%%%%%%%%%%%%%%%%%%

%%%%%%%%%%%%%%%%%%%%%%%%%%%%%%%%%%%%%%%%%%%%%%%%%%%%%%%%%%%%%%%%
\section*{Appendix A. Properties of the Residual Squared Error }
\label{app:emin}
%%%%%%%%%%%%%%%%%%%%%%%%%%%%%%%%%%%%%%%%%%%%%%%%%%%%%%%%%%%%%%%%

We noted at the beginning of Section \ref{sec:LSM} that if the vectors
\phii are mutually orthonormal, then the optimal measurement is a
set of projections onto the states \phii, and the resulting squared error is
zero. In this case $S=\Phi^* \Phi=I_m$, and $\sigma_i=1,\,\,1 \leq i\leq m$. 

If the vectors \phii are normalized but not orthogonal, then
we may decompose $S$ as $S=I_m+D$, where $D$ is the matrix of inner products
$\braket{\phi_i}{\phi_j}$ for $i \neq j$ and has diagonal elements all
equal to $0$. We expect that if the inner products are relatively
small, \ie if the states \phii are nearly orthonormal, then we will be able
to distinguish between them pretty well; equivalently, we would expect
the singular values to be close to $1$. Indeed, from \cite{Horn} we have the
following bound on the singular values of $S=I+D$:
\begin{equation}
|\sigma_i^2-1|^2 \leq \tr(D^*D),\,\,\,1 \leq i \leq m.
\end{equation}

We now point out some properties of the  minimal
achievable squared error $E_{min}$ given by  (\ref{eq:emin2}).
For a given $m$, $E_{min}$ depends only on the  singular values of the
matrix $\Phi$. Consequently, 
any linear operation on the vectors \phii that does not affect the
singular values of $\Phi$ will not affect $E_{min}$. 

For example, if we obtain a  new set of states
$\ket{\phi'_i}$ by unitary mixing of the states \phii, \ie
$\Phi'=\Phi Q^*$ where $Q$ is an $m \times m$ unitary
matrix, then the new optimal measurement vectors
$\ket{\mu'_i}$ will typically differ from the measurement vectors
$\ket{\hat{\mu}_i}$; however
the minimal achievable squared error is the same. Indeed, 
defining $S'=\Phi'^*\Phi'=Q S Q^*$, where
$S=\Phi^*\Phi$, we see that the matrices $S'$ and $S$ are
related through a similarity transformation 
and consequently have equal \mbox{eigenvalues \cite{Horn}}. 

Next, suppose we obtain a new set of states
$\ket{\phi'_i}$ by a general nonsingular linear mixing of the
states $\ket{\phi_i}$, \ie $\Phi'=\Phi A^*$,
where $A$ is an arbitrary $m \times m$
nonsingular matrix. In this case the eigenvalues of 
$S'=A S A^*$ will in general differ from the eigenvalues of $S$.
Nevertheless, we have the following theorem:
\begin{theorem}
\label{thm:emin}
Let $E_{min}$ and $E'_{min}$ denote the minimal achievable squared
error when distinguishing between the pure state ensembles
$\{ \ket{\phi_i} \}$
and $\{\ket{\phi'_i}\}$ respectively, where
\mbox{$\ket{\phi'_i}=\sum_{j=1}^m a_{ij}^* \ket{\phi_j}$}.
Let $A$ denote the matrix whose 
$ij$th element is $a_{ij}$. Let
$\lambda_1(AA^*)$ and $\lambda_m(AA^*)$ denote the largest and smallest
eigenvalues of $AA^*$ respectively, and let $\{\sigma_i,\,1 \leq i
\leq r\}$ denote the singular values of the matrix $\Phi$ of columns
\phii. 
Then, 
\[  2\bl 1-\sqrt{\lambda_1(AA^*)}\br\sum_{i=1}^r \sigma_i \leq 
E'_{min}-E_{min} \leq 
2 \bl 1-\sqrt{\lambda_m(AA^*)} \br \sum_{i=1}^r \sigma_i.\]
Thus, $E'_{min} \leq E_{min}$ if $\lambda_m(AA^*) \geq1$ and
$E'_{min}\geq E_{min}$ if $\lambda_1(AA^*)\leq 1$. \\
In particular, if $A$ is unitary then $E_{min}=E'_{min}$.
\end{theorem}

{\em Proof:} 
We rely on the following theorem due to Ostrowski (see \eg \cite{Horn},
p.~224): 

{\em Ostrowski Theorem:}
Let $A$ and $S$ denote $m \times m$ matrices with $S$ Hermitian and $A$
nonsingular, and let $S'=ASA^*$.
Let $\lambda_k(\cdot)$ denote the $k$th eigenvalue of the 
corresponding matrix, where the eigenvalues are arranged in decreasing
order. For every $1 \leq i \leq m$, there exists a positive real
number $a_i$ such that $\lambda_m(AA^*) \leq a_i \leq \lambda_1(AA^*)$
and $\lambda_i(S')= a_i \lambda_i(S)$.

Combining this theorem with the expression (\ref{eq:emin2})
for the residual squared error results in 
$E'_{min}-E_{min}=2 \sum_{i=1}^r \bl 1-\sqrt{a_i} \br \sigma_i$.
Substituting $\lambda_m(AA^*) \leq a_i \leq \lambda_1(AA^*)$  results
in Theorem \ref{thm:emin}. If $A$ is unitary, then $AA^*=I$, and
$\lambda_i(AA^*)=1$ for all $i$. 

%\bibliography{srm}
%\bibliographystyle{plain}

\newpage
%%%%%%%%%%%%%%%%%%%%%%%%%%%%%%%%

\newpage 

\setlength{\unitlength}{.2in}
\begin{figure}[h]
\begin{center}
\begin{picture}(15,14)(0,0)
%\put(0,0){\framebox(15,14){}}
\put(7,6){\vector(0,1){7}}
\put(7,6){\line(0,-1){4}}
\put(7,6){\line(-1,0){6}}
\put(7,6){\line(-3,5){3}}
\put(4.06,10.9){\vector(-1,1){.2}}
\put(3.6,12){\makebox(0,0){\small $\ket{\phi_2}$}}
\put(7,6){\vector(1,0){6}}
\put(14.3,6){\makebox(0,0){\small $\ket{\phi_1}$}}
\put(7,6){\vector(4,1){5.82}}
\put(14,7.9){\makebox(0,0){\small $\ket{\hat{\mu}_1}$}}
\put(7,6){\vector(-1,4){1.45}}
\put(5.6,12.5){\makebox(0,0){\small $\ket{\hat{\mu}_2}$}}

\dashline{.2}(13,6)(12.8,7.4)
\put(12.93,6.3){\vector(1,-3){.1}}
\put(12.2,6.65){\makebox(0,0){\footnotesize $\ket{e_1}$}}
\dashline{.2}(5.58,11.8)(4,11.1)
\put(4,11.1){\vector(-1,-1){.1}}
\put(5.1,10.85){\makebox(0,0){\footnotesize $\ket{e_2}$}}

\end{picture}
\caption{$2$-dimensional example of the LSM. The state vectors
$\ket{\phi_1}$ and $\ket{\phi_2}$ are given by (\ref{eq:phiex}), the
optimal measurement vectors $\ket{\hat{\mu}_1}$ and
$\ket{\hat{\mu}_2}$ are given by (\ref{eq:muex}) and are orthonormal, and
$\ket{e_1}$ and $\ket{e_2}$ denote the error vectors defined in
(\ref{eq:error}). } 
\label{fig:srmex}
\end{center}
\end{figure}
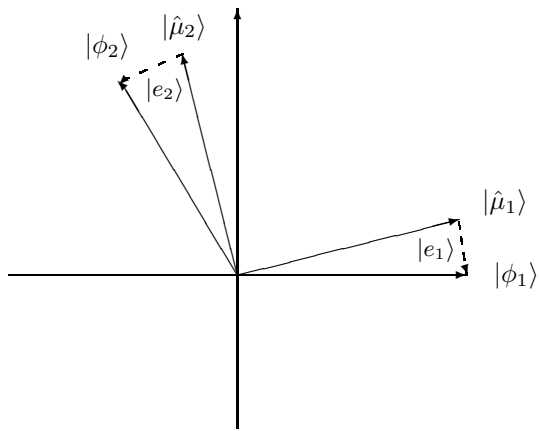
\begin{figure}[hb]
\begin{center}
\resizebox{7cm}{5cm} 
{\includegraphics{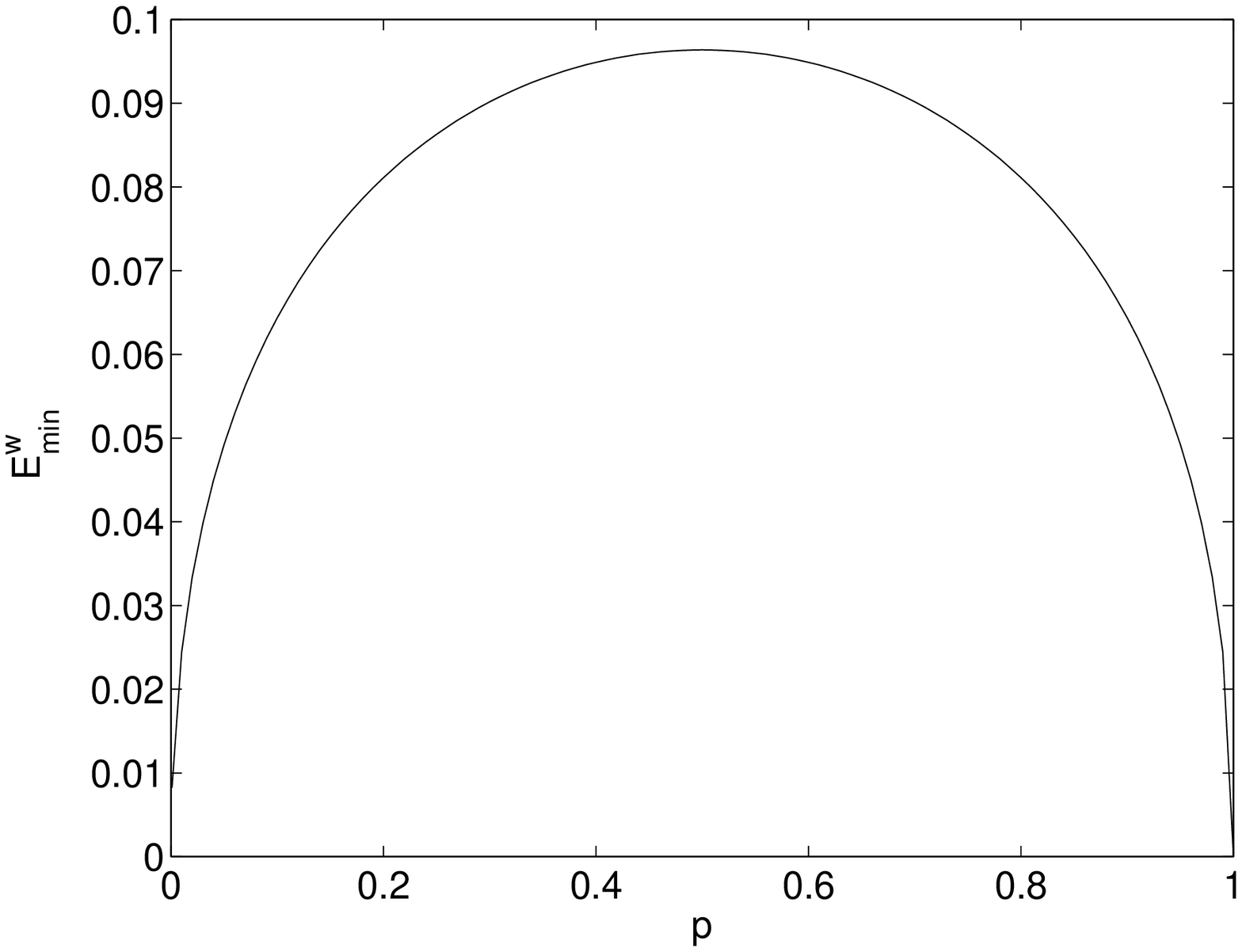}}
\end{center}
\caption{Residual squared error $E_{min}^w$ (\ref{eq:mwerr}) as a
function of $p$, the prior probability of $\ket{\phi_1}$, when using a WLSM.
The weights are chosen as $w_1=\sqrt{p}$ and $w_2=\sqrt{1-p}$. 
For $p=1/2$ the WLSM and the LSM coincide. }
\label{fig:wlsmex}
\end{figure}

\setlength{\unitlength}{.2in}
\begin{figure}[h]
\begin{center}
\begin{picture}(15,14)(0,0)
%\put(0,0){\framebox(15,14){}}
\put(7,6){\vector(0,1){7}}
\put(7,6){\line(0,-1){4}}
\put(7,6){\line(-1,0){6}}
\put(7,6){\line(-3,5){3}}
\put(4.06,10.9){\vector(-1,1){.2}}
\put(3.6,12){\makebox(0,0){\small $\ket{\phi_2}$}}
\put(7,6){\vector(1,0){6}}
\put(14.3,6){\makebox(0,0){\small $\ket{\phi_1}$}}
\put(7,6){\vector(4,1){5.82}}
\put(14,7.9){\makebox(0,0){\small $\ket{\hat{\mu}_1}$}}
\put(7,6){\vector(-1,4){1.45}}
\put(5.6,12.5){\makebox(0,0){\small $\ket{\hat{\mu}_2}$}}

\dashline{.2}(7,6)(10,11)
\dashline{.2}(7,6)(5,2.3)

\end{picture}
\caption{Symmetry property of the state set
$\SSS=\{\ket{\phi_1},\ket{\phi_2}\}$ and the optimum measurement
vectors $\{\ket{\hat{\mu}_1},\ket{\hat{\mu}_2}\}$.
$\ket{\phi_1}$ and $\ket{\phi_2}$ are given by (\ref{eq:phiex}), and
$\ket{\hat{\mu}_1}$ and
$\ket{\hat{\mu}_2}$ are given by (\ref{eq:muex}). 
Because the state vectors
are invariant under a reflection about the dashed line, the optimum
measurement vectors must also have this property.  In addition, the
measurement vectors must be orthonormal.
The symmetry and orthonormality
properties completely determine the optimum measurement vectors
$\{\ket{\hat{\mu}_1}, \ket{\hat{\mu}_2}\}$ (up to sign reversal).}
\label{fig:guex}
\end{center}
\end{figure}

\clearpage 
\listoffigures

\end{document}